# Nanoscale Piezoelectric Response across a Single Antiparallel Ferroelectric Domain Wall

David A Scrymgeour and Venkatraman Gopalan

Department of Materials Science & Engineering and Materials Research Institute

Penn State University

Surprising asymmetry in the local electromechanical response across a single antiparallel ferroelectric domain wall is reported. Piezoelectric force microscopy is used to investigate both the in-plane and out-of- plane electromechanical signals around domain walls in congruent and near-stoichiometric lithium niobate. The observed asymmetry is shown to have a strong correlation to crystal stoichiometry, suggesting defect-domain wall interactions. A defect-dipole model is proposed. Finite element method is used to simulate the electromechanical processes at the wall and reconstruct the images. For the near-stoichiometric composition, good agreement is found in both form and magnitude. Some discrepancy remains between the experimental and modeling widths of the imaged effects across a wall. This is analyzed from the perspective of possible electrostatic contributions to the imaging process, as well as local changes in the material properties in the vicinity of the wall.

# I. Introduction

In a uniaxial ferroelectric, two ferroelectric domain orientations are possible: along the uniaxial +*c* axis and the –*c* axis. A 180° domain wall separates these two domain states. By controlling the orientation of the domain structures, many devices can be fabricated in ferroelectrics such as lithium niobate, $LiNbO_3$, and lithium tantalate, $LiTaO_3$. Of these, the most common is quasi-phase matched second harmonic generation where the period of the domain grating structure determines the frequency of input light that is most efficiently frequency converted.[1] Other devices based on domain patterning include electro-optic gratings, lenses, and scanners, which require manipulation of the domain shapes into more intricate geometries.[2-4] These applications, among others, exploit the fact that antiparallel domains have identical magnitudes, but differ in the sign of the odd-rank coefficients of piezoelectric, ($d_{ijk}$), electro-optic,($r_{ijk}$) and third-rank nonlinear optical ($d_{ijk}$) tensors, where dummy subscripts refer to crystal physics axes in an orthogonal coordinate system. The second rank properties such as refractive indices are expected to be identical across a domain wall.

The local nature of antiparallel domain walls is a fundamental property of interest. However, recent studies on $LiNbO_3$ and $LiTaO_3$ suggest that antiparallel domain walls can exist with differing refractive indices and lattice parameters across a 180° wall.[5] Such asymmetry in optical and elastic properties across a wall is unexpected and has been shown to arise from the presence of non-stoichiometric defects in these crystals.[6] Here we show that local electromechanical properties across these walls in lithium niobate show an asymmetric response as well. We present a detailed experimental and theoretical modeling investigation of the piezoelectric response at a single antiparallel ferroelectric domain wall. This is probed using a scanning probe microscopy technique called piezoelectric force microscopy (PFM). Together, these results suggest that while the structure of an ideal ferroelectric domain wall is well understood to be atomically sharp (1 to 2 unit cells wide)[7] small amounts of defects can change the local structure of a domain wall dramatically through defect-domain wall interactions.

This paper is organized as follows. The defect-domain wall interactions in $LiNbO_3$ are described in Section II. Section III presents the PFM results. Section IV describes the theoretical modeling of the observed piezoelectric response at the walls. Finally, a comparison between experiments and modeling is presented, and the results discussed in Section V.

## II. DOMAIN WALLS AND STOICHIOMETRY IN $LiNbO_3$

Stoichiometric $LiNbO_3$ has a composition ratio of C = [Li]/[Li+Nb] = [Nb]/[Li+Nb] = 0.5. However, commercially available congruent lithium niobate, denoted by $(Li_{0.95}Nb_{0.0}\square_{0.04})NbO_3$, is lithium deficient with composition ratio C = [Li]/[Li+Nb] = 0.485. This leads to nonstoichiometric defects, which are presently believed to be Nb-antisites, $Nb_{Li}$, (which are excess Nb atoms at Li locations), and lithium vacancies denoted by $\square_{Li}$.[8] The defect equilibrium is $4[Nb_{Li}]=[\square_{Li}]$.

These point defects give rise to an order of magnitude increase in the coercive field, a large internal field, and the presence of local structure at domain walls in the congruent crystal composition.[6]

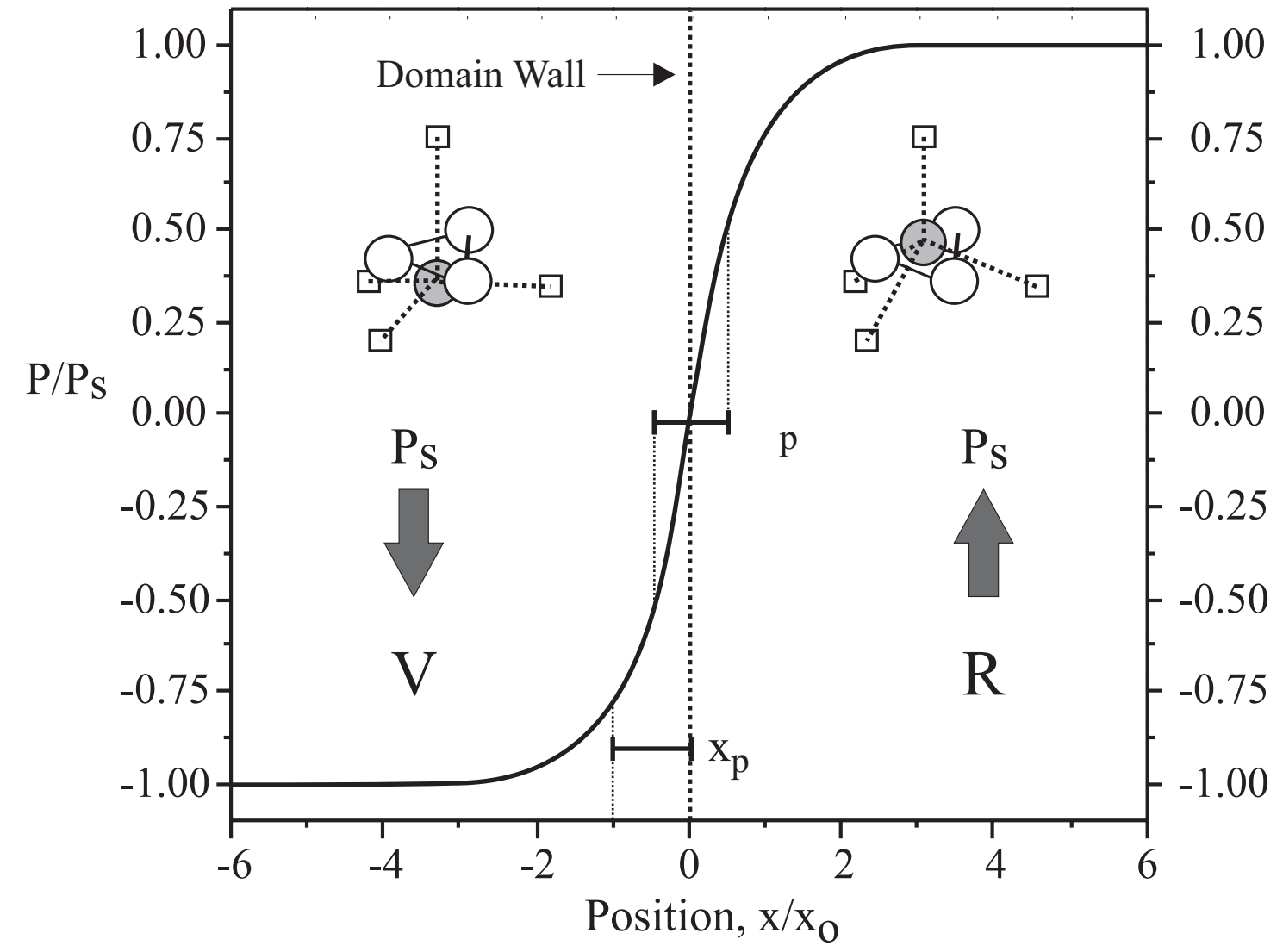

**Figure 1: The variation of the normalized polarization, $P/P_s=tanh(x/x_p)$, across a single 180° ferroelectric domain wall and a schematic of nonstoichiometric defect dipoles in congruent lithium niobate. Open circles indicate oxygen atoms, the filled circle is the Nb-antisite defect, and the square symbols are lithium vacancies. The virgin (V) state contains stable defects and the domain reversed state (R) created at room temperature has frustrated defect dipoles. The full width at half maximum is denoted $\omega_p$.**

As proposed by Kim[6], these defects are not random, but can possess a low energy configuration, called a defect dipole, such as shown in Figure 1 schematically. In a crystal grown from high temperature, all the defect dipoles have the low energy configuration, and the domain state is labeled "virgin state" (labeled hereafter as V). When the domain is reversed at room temperature using electric fields, domains and domain walls are created, which are in "domain reversed state" (labeled hereafter as R). Within these domains, the defects are in the "frustrated state" wherein the Nb atom has moved, but the lithium vacancies are "stuck" in a frustrated state, unable to move due to negligible ionic conductivity at room temperature. A domain wall at room temperature between a virgin state and a reverse state therefore represents not only a transition of the lattice polarization, $P_s$ from an up- to a down- state, but also from a stable to a frustrated defect state, respectively. The transition of the lattice polarization from an up to a down state is given by $P = P_s \tanh(x/x_p)$ where $x_p$ is the half width of the wall.[9] While the lattice polarization may indeed switch over a few unit cells, the transition of defect states across a wall appears to give rise to broad index and strain change in the wall region[10]. However, it has been shown that by annealing such a crystal at >150°C, this defect frustration is considerably relieved.[11]

In this paper the interaction of these nonstoichiometric point defects with the domain wall will be examined through the measurement of electromechanical properties. Crystals of congruent and near-stoichiometric compositions (C=0.499) of $LiNbO_3$ will be compared. We note that *near-stoichiometric* crystals are still not perfectly stoichiometric crystals, and still exhibit small defect influences on the domain reversal properties, such as an internal field of ~0.1 kV/mm, in comparison to internal fields of ~3 kV/mm in congruent $LiNbO_3$. A detailed modeling of the piezoelectric force microscopy images will also be presented *for the near-stoichiometric compositions*.

## III. PIEZOELECTRIC FORCE MICROSCOPY: EXPERIMENTS

### A. Principle of Operation

The use of scanning probe techniques in the investigation of ferroelectric domain structure are well established.[12-15] PFM especially has been used to study the antiparallel domain states of bulk crystals like triglycine sulfate (TGS) [16, 17] and thin film piezoelectric samples of random domain orientation.[18, 19] As shown in Figure 2, the technique involves bringing a conductive tip in contact with the sample surface, a distance of 0.1 to 1 nm. A modulated AC voltage is applied to the sample through the tip, and the first harmonic oscillations of the cantilever are detected by a lock-in technique. If the sample surface is piezoelectric, the oscillating electric field causes deflection of the sample surface through the converse piezoelectric effect. This offers a technique to examine domain and domain structures of piezoelectric materials at the micrometer and nanometer scale, as the electromechanical response of the sample surface gives information about the orientation of the polarization direction below the tip as well as the relative orientations between adjacent domains.

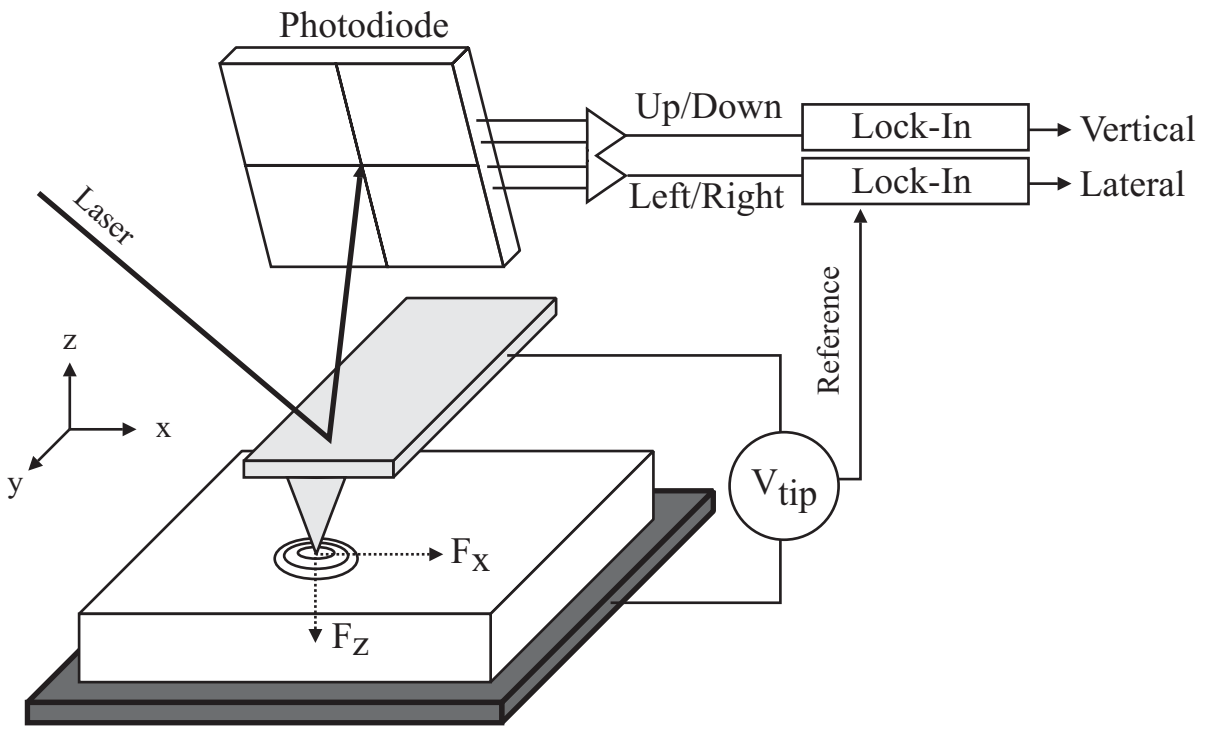

**Figure 2: Schematic of piezoelectric force microscopy (PFM) setup. The forces acting in the vertical plane ($F_z$) give the vertical signal, the forces in the horizontal plane ($F_x$) gives the lateral signal. $V_{tip}$ is an oscillating voltage applied to the sample. Up and down are the signals from the top and bottom 2 quadrants of the photodiode, while left and right are the signals from the left and right 2 quadrants.**

If an oscillating voltage of the form

$$V_{tip} = V_{dc} + V_{ac}cos(\omega t) \quad \text{Equation 1}$$

is applied to a piezoelectric surface, the amplitude of displacement of the surface is given by

$$d = d_0 + A\cos(\omega t + \varphi) \qquad \text{Equation 2}$$

where $d$ is deformation of the sample surface, $d_0$ is the static deflection due to any bias voltage, $A$ is the amplitude of the oscillation, $\omega$ is the frequency of oscillation, and $\varphi$ is the phase of the electromechanical response of the sample surface. In the *vertical imaging* mode, surface displacements perpendicular to the sample surface, both the amplitude (the magnitude of surface displacement) and the phase (a measurement of the phase delay between the applied electric field and the response of the sample surface), are measured.[12, 20, 21] In-plane oscillations can also be investigated by observing the torsioning of the cantilever in the *lateral imaging* mode.[22]

**B. Samples and Measurement Details**

Z-cut Lithium niobate crystals (polarization, $P_s$, along the thickness direction) with thickness ~300 μm were used in this study. Randomly nucleated antiparallel domains were created in the crystals by electric field poling starting from a single domain state. Briefly, two water cells located on the opposite sides of the crystal were used as electrodes. Electric fields greater than the coercive field of the crystal, (~22 kV/mm in these crystals), were applied by applying slowly ramping voltage to the water cells at room temperature. At the onset of nucleation of domain shapes, the field was removed when the domain poling process was partially completed leaving many small domains of opposite orientation in a matrix of original orientation. The domain sizes created varied from 4 to 500 μm with average size of ~100 μm.

Measurements were made using an Explorer AFM head manufactured by Thermomicroscopes. Cantilevers (fabricated by Micromasch) of varying stiffness from 2 and 20 N/m were used in the imaging. The tips were coated with Ti-Pt and are electrically connected to an external voltage supply with the ground plane on the back of the sample mount. Since coated tips degrade due to the peel-off of the conductive coating, the tips were replaced frequently and images presented in this paper where taken with minimally used tips (only enough to characterize the tips and locate the feature of interest). The tips have a nominal radius of curvature of 50 nm as provided by the vendor, but the exact radius of curvature will be slightly different dependent upon the degree of use. A Stanford system SR830 lock-in amplifier was used to lock onto the raw CCD signals and to generate the imaging oscillation voltage waveform. A HP32120 function generator was sometimes used to generate higher voltage signals (up to 10 V peak). Most images were taken at 5 V peak (3.5 RMS) imaging voltage with the frequency of oscillation around 35 kHz.

Our system was calibrated using a similar technique to Christman[23], where the amplitude of a uniformly electroded sample of *x*-cut quartz was measured as a function of applied voltage for various low frequency oscillations and contact forces. The slope of the maximum amplitude versus applied voltage of the sample surface was assumed to be equal to the $d_{11}$ coefficient of the quartz at low frequencies of 1 kHz or less. This allowed us to calibrate the amplitude of surface displacement (measured as an amplitude signal on the lock in amplifier) and a physical displacement of the surface at a particular frequency.

In general, the frequency of the oscillating probe voltage at the tip plays a very important role in determining the amplitude and contrast of measurements in PFM. Choosing the proper frequency can enhance or minimize contrast in the image, or even null a contrast completely. Labardi examined the influence of frequency on the measurement technique, attributing most of the variation in oscillation to a complex resonant structure determined by the tip and sample surface in contact with each other.[24, 25]

Frequency scans of the sample were made by keeping the probe over a uniform domain area in lithium niobate and varying the frequency of the applied voltage and plotting the resulting cantilever amplitude and the phase between applied voltage and surface response. The phase shows a continual increase in angle which indicates a frequency dependent background term.[25] Images in this paper were taken around 35 kHz at a relatively flat area in the amplitude and phase

There are several different origins of the signal in a vertical PFM image. The net amplitude, $A$, of the oscillating surface is given by the sum of all the contributing factors

$$A = A_{pi} + A_{es} + A_{nl} \qquad \text{Equation 3}$$

where $A_{pi}$ is the electromechanical (piezoelectric) amplitude, $A_{es}$ is the electrostatic amplitude,[20, 26] and $A_{nl}$ is the non-local contribution due to capacitive interaction between the sample surface and the cantilever assembly.[27] Discussion of the magnitudes of each factor in Equation 3 are discussed in detail in papers by Hong[28] and Kalinin.[29] Any signal observed on a sample, then, must be thought of as the sum of all these interactions.

The mechanism for piezoelectric signal is as shown in Figure 3(a). Here, the amplitude of the vertical oscillations ($A_{pi}$) should be the same on either side of the domain wall and be related to the piezoelectric coefficient $d_{33}$. The phase, which is the delay between the applied signal and the surface displacement, contains information about the polarization direction. For example, in lithium niobate, where the piezoelectric $d_{33}$ coefficient is positive, the application of positive tip bias to the $+P_s$ surface of a domain (i.e. positive end of the polarization, $P_s$), results in a contraction of the sample surface (negative displacement of cantilever, $-A_{pi}$) as shown in Figure 3(a). Therefore, the surface oscillation is π out of phase with the oscillating tip bias. The case is reversed above a $-P_s$ surface of a domain (i.e. negative end of the polarization, $P_s$) in Figure 3(b) where the surface oscillation is in phase with the oscillating tip bias.

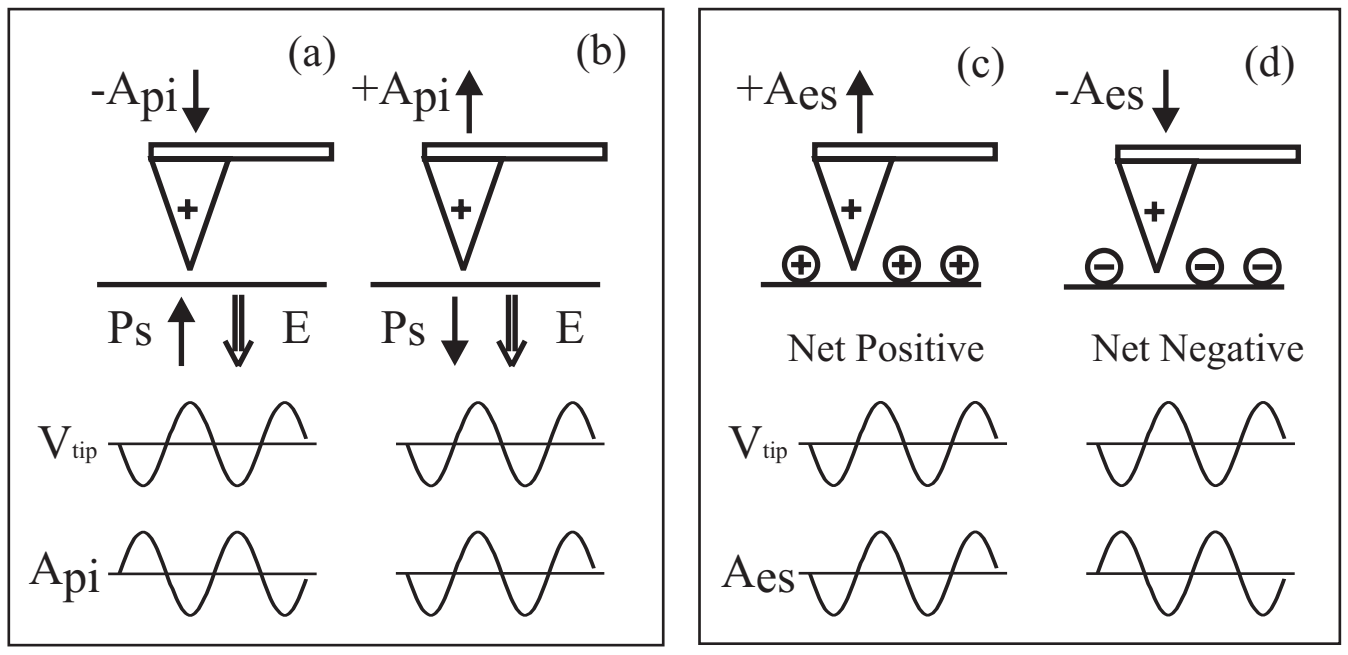

**Figure 3: Mechanism of contrast for piezoelectric signal (a,b) and electrostatic signal (c,d) where $V_{tip}$ is the oscillating voltage applied to the sample, $A_{pi}$ is the piezoelectric amplitude, and $A_{es}$ is the electrostatic amplitude. Down arrows indicate negative amplitude, $-A_{pi}$ and $–A_{es}$, up arrows indicate positive amplitude, $+A_{pi}$ and $+A_{es}$.**

The electrostatic response, also called the "Maxwell stress", are electrostatic forces acting between a conductive cantilever tip and a charged ferroelectric surface as shown in Figure 3(c,d). The net charge of the surface, due to screening of the spontaneous polarization, can be either net positive or negative. For a partially screened surface, a net positive charge will be on the $+P_s$ surface and net negative charge on the $-P_s$ surface. In the case of an over-screened surface, where the spontaneous polarization is over-compensated by surface charges, a net negative charge will be on the $+P_s$ surface and net positive charge on the $-P_s$ surface. Partially or completely screened surfaces are the likely state of ferroelectric surfaces in air[30], while over screening is observed on electrically poled samples like PZT thin films.[31, 32] We note that the phase relation between the applied voltage and the tip displacement is the same for piezoelectric mechanism and the over-compensated surface. The phase relation is opposite for the partially screened case.

On a piezoelectric surface, all contrast mechanisms are active. To test for the dominant mechanism for a given sample, the relative phase delay above a domain of known orientation must be found. Using a lock-in amplifier, we have experimentally verified that above $+P_s$ surface in lithium niobate, the oscillation of the sample is phase shifted 180º from the input oscillating voltage and in-phase above a $-P_s$ surface. This indicates two possibilities: (1) the signal is primarily electromechanical in nature or (2) the $-P_s$ surface has a net negative charge and the $+P_s$ surface has a net positive charge, which indicates an over-screened surface. Both of these contributions could be occurring simultaneously and will be analyzed in the discussion section. As pointed out before, the frequency of the imaging voltage can cause a large variation in signal amplitude and phase, so care must be taken to avoid frequencies close to resonant peaks or the dominant mechanism can be incorrectly identified.

In addition to the local tip-surface interactions, there is also a long range electrostatic interaction due to capacitive cantilever assembly-surface interactions, $A_{nl}$. If this interaction is strong enough it can obscure important image characteristics, like the phase shift between adjacent domains.[27] It depends inversely on the spring constant of the cantilever and can be minimized by using very stiff spring constant cantilevers.[28] Measurements were made with cantilevers of stiffness varying between 2 and 20 N/m. It was found that for stiffness less than ~12 N/m a proper 180° phase shift between adjacent domains could not be seen regardless of the imaging frequency. All images in this paper were taken with cantilevers of spring constant 14 N/m.

**C. Vertical Imaging Mode Piezoelectric Response**

PFM images a variety of interactions at the domain wall - mechanical, electro-mechanical, and electro-static. Therefore, the wall width found in PFM images, as determined by the amplitude, is actually the *interaction width*, which we note, is *not* to be confused with the explicit domain wall width over which the polarization reverses. The latter has been measured by Bursill to have an upper limit of 0.28 nm using high-resolution TEM images in lithium tantalate (isomorphous to lithium niobate).[7]

The interaction widths, $\omega_o$, of all images presented in this paper are defined as the full width at half maximum (FWHM) corresponding to the amplitude change from the minimum point to where the value increased to half the full value on either side. We should note that the FWHM, $\omega_o$, is different than $x_o$ used in the expression $A = A_o \tanh(x/x_o)$. At $\pm\omega_o/2$ the amplitude is $A = \pm 0.5\ A_o$, compared to positions at $\pm x_o$ where the amplitude $A = \pm 0.76\ A_o$. This is shown graphically in Figure 1. For a symmetric curve, the interaction width (FWHM), $\omega_o$, is related to the half wall width, $x_o$, as $x_o = 0.91\omega_o$.

V. Bermudez imaged Czochralski grown periodically poled congruent lithium niobate with a variety of techniques including PFM [33]. They did not measure the interaction width of the domain wall but rather measured the amplitude of oscillation in a uniform domain area. J. Wittborn imaged room temperature periodically poled lithium niobate and determined the interaction width (FWHM) to be ~150 nm.[34] Gruverman measured the interaction width in electrically poled domains in lithium tantalate to be 120 nm[35]. However, since the signal used to determine the interaction width (amplitude, phase, or X=amplitude×cos(phase) signal) or the frequency of the applied voltage were not mentioned comparison of these papers with the current work is limited.

Shown in Figure 4 are the topography, amplitude and phase images of a region containing a domain wall in congruent $LiNbO_3$. A topographic step across the domain wall was not measured on any crystal, which is attributed to the presence of residual polishing scratches of approximately 2-3 nm visible in Figure 4(a). Non-local electrostatic interaction in the image has been minimized as evidenced from the similar vertical displacement amplitude on either side of the wall (Figure 4(e)) and a proper 180° phase change across the wall (Figure 4(f)). There is very little cross talk between the topography image and the PFM image.

Measurements were then taken of the interaction widths in unannealed congruent crystals. After a domain wall was located, consecutive images were taken on the same area, zooming in on the domain wall. The time constant on the lock-in amplifier was made as small as possible (30 μs) and scans were taken very slowly (scan rates < 2000 nm/s) to achieve the highest resolution of the interaction width.

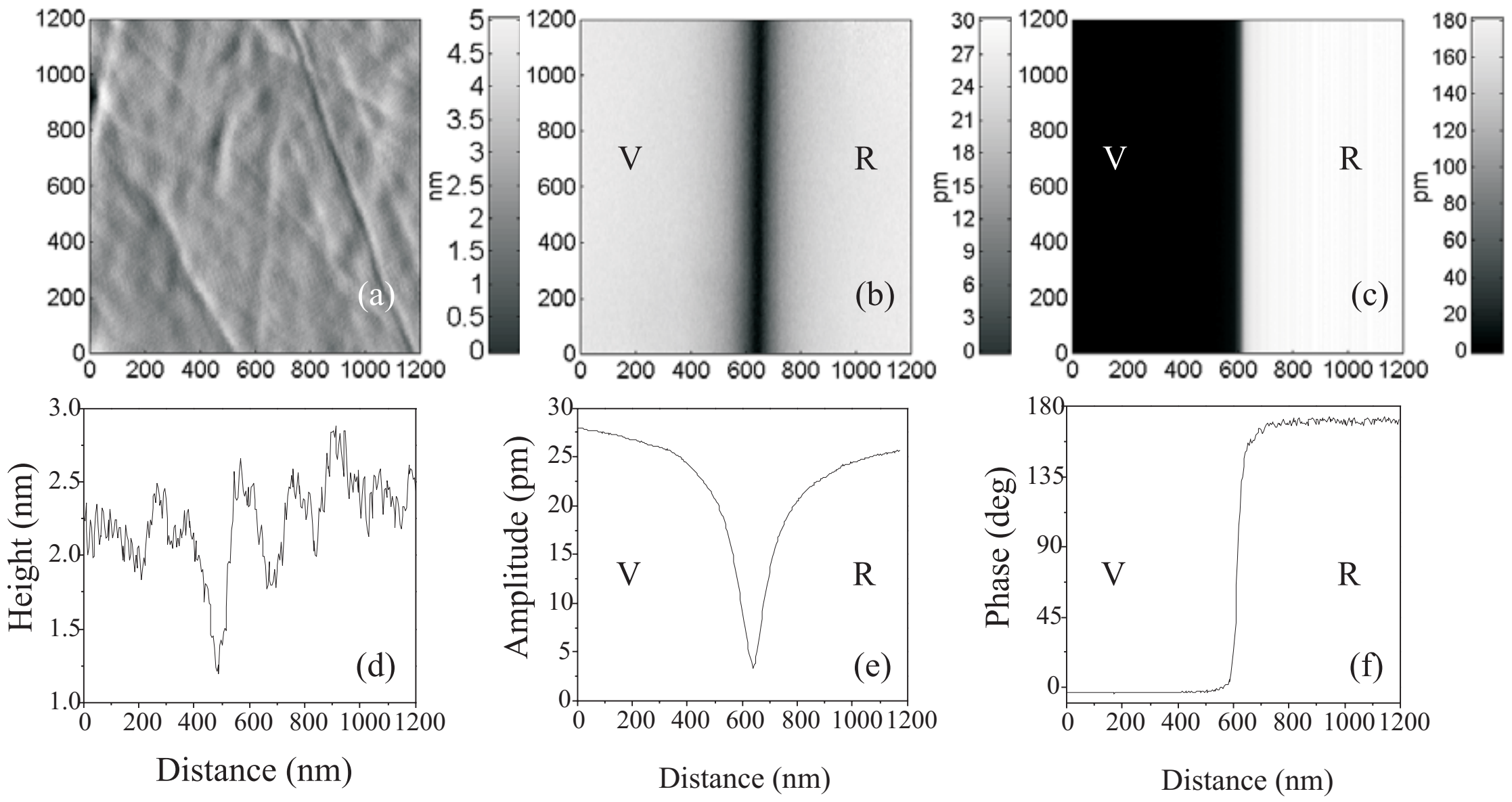

**Figure 4: Images on congruent lithium niobate. (a), (d) are topography images and a cross section; (b),(e) are vertical amplitude and cross section, and (c),(f) are phase image and cross section, respectively. V is the virgin side; R is the domain-reversed area. Distances in (a), (b), and (c) are in nanometers.**

Close analysis of the vertical amplitude PFM signal scans of the congruent domain wall shows an asymmetry as shown in Figure 4(e). The long tail region in the signal is always present on the domain-reversed side (R) which is created by electric fields at room temperature and contains frustrated defect dipoles. Scan artifacts have been eliminated as a source of asymmetry by comparing images obtained by scanning in both forward and reverse directions as well as scanning with the cantilever perpendicular (0°) and parallel (90°) to the domain wall. The asymmetry is still present. To eliminate leveling or background artifacts, several correction functions have been applied to the profiles. For example, using a hyperbolic tangent correction to mimic leveling artifacts leaves the asymmetric profile unchanged.

Since this asymmetry is not an artifact of leveling or scanning, it indicates the presence of local structure around the domain wall. This asymmetry could be related to the intrinsic nonstoichiometric defects present in the material. This is further supported by a comparison of near-stoichiometric lithium niobate crystals to congruent crystals as shown in Figure 5(a). The asymmetry is almost completely absent in the near-stoichiometric crystals. The interaction widths were found from the amplitude images of several samples and different domain sizes. Since the amplitudes were often not the same on either side of the wall, the interaction width, $\omega_o$, was found from the minimum point to where the value increased to half the full value on either side. The smallest interaction width in congruent lithium niobate was ~140 nm, and in near-stoichiometric lithium niobate, ~ 113 nm. The near-stoichiometric crystal width is ~20% less than the congruent crystal indicating the influence of defect dipoles.

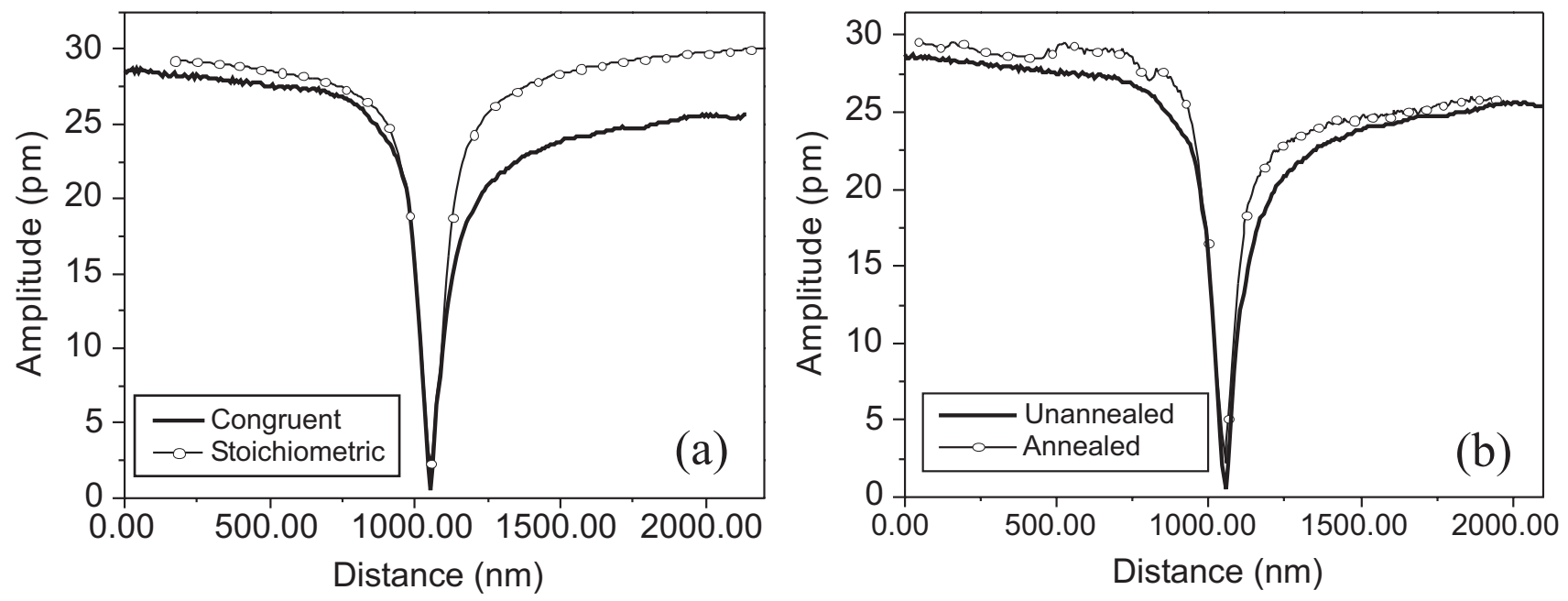

**Figure 5: Effects of nonstoichiometry on vertical PFM signal. (a) Comparison of congruent and near-stoichiometric lithium niobate vertical amplitude images.**

**Notice the asymmetry in congruent case. (b) Comparison of annealed and unannealed crystals in congruent crystals.**

Further support for the role of nonstoichiometric point defects as the origin of the asymmetry is obtained by comparing measurements taken before and after annealing of the congruent crystals at 200°C for 24 hours. This anneal allows for the reorientation of the frustrated defect dipoles in the domain state R. Looking at the same domain wall, the interaction width is found to decrease slightly as shown in Figure 5(b), reducing from ~140 nm in the unannealed crystal to ~120 nm in the annealed crystal.

The asymmetry could also be related to a mechanical clamping of the inner domain, as it is effectively embedded in a matrix of oppositely oriented domain. However, this has been eliminated as a possibility, by examining many walls of domains of varying sizes. Even in very large domains sizes, such as a 4 mm domain in a sample of 10 mm, the asymmetry was still present.

Changes in the sample surface properties, such as local conductivity, could also give rise to the sample asymmetry. However, this is unlikely because $LiNbO_3$ is inherently non-conducting at room temperature, with an energy barrier of 1.1 eV for hopping conduction and room temperature conductivity as $10^{-18}$ Ω cm.[36] Studies of ferroelectric oxide surfaces do not consider conductivity to be a major factor in imaging contrast across a domain wall.[30, 37]

### D. Lateral Imaging Mode Piezoelectric Response

Lithium niobate belongs to point group *3m*, and the domains form with the crystallographic *y*-directions parallel to the domain walls as shown in Figure 6(a). The lateral image can then probe information in two different planes. When the cantilever arm is parallel to the domain wall as shown in Figure 6(b) and (d), the distortions in the crystallographic *x-z* plane are probed. This will be referred to as a *0° lateral scan* for the remainder of this paper. On crossing from one domain orientation to the other across the domain wall in the *x-z* plane, the *z-* and *y-* crystallographic axes changes direction (from –*z* (-*y*) to +*z* (+*y*)) through a two-fold rotation about the *x*-axis. Wittorn [34] proposed that the contrast comes mainly from a distortion of the sample surface near a domain wall as one side expands up and the other shrinks down, giving a sloping surface at the domain wall as pictured in Figure 6(b). In this case, only the domain wall region will show a maximum in the lateral signal.

When the cantilever arm is perpendicular to the wall as shown in Figure 6(c,e), distortions and torsions in the *y-z* plane are probed. This will be referred to as a *90° lateral scan* for the remainder of this paper.

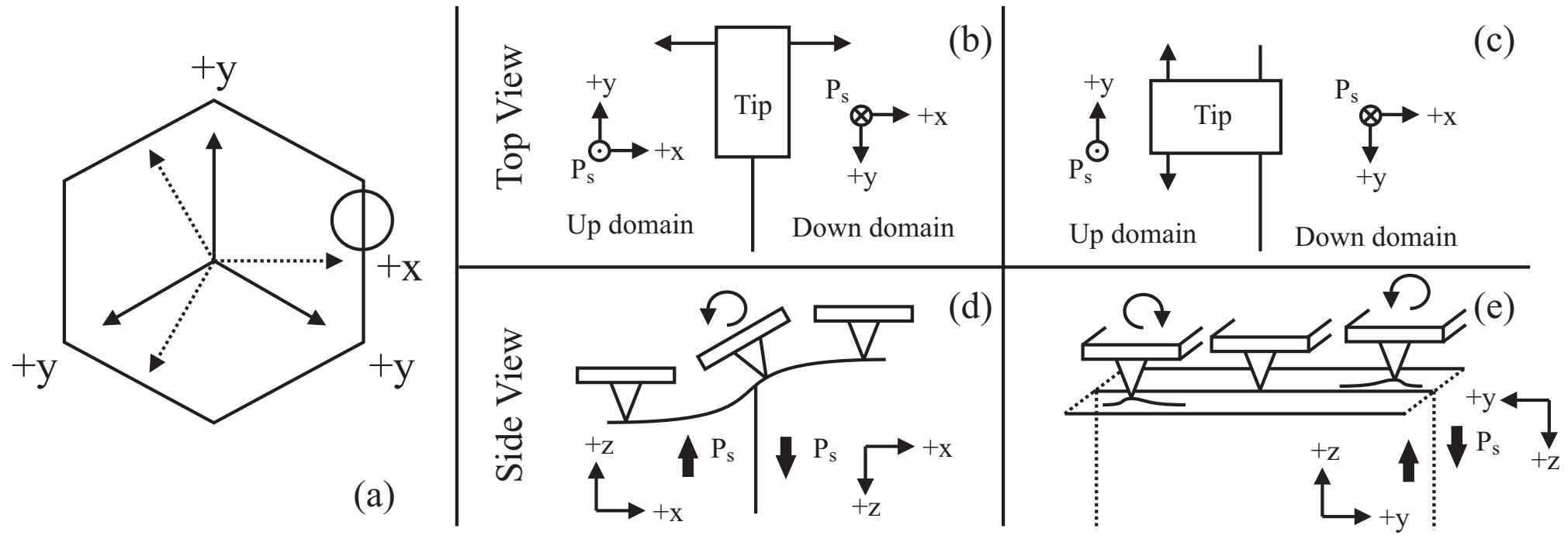

**Figure 6: The importance of symmetry in lateral images in $LiNbO_3$. (a) The domain structure relative to the *x-y* crystallographic axes. The circled area is**

**expanded in (b-e). Cantilever parallel to domain wall is shown in top view (b) and side view (d). Scanning is in the horizontal direction shown by arrows. Cantilever perpendicular to domain wall is shown in top view (c) and side view (e) scanning in vertical direction shown by arrows. Loops in indicate torsion on cantilever.**

The profiles with the cantilever parallel to the domain wall (0° scan) are shown in Figure 7. They indeed show a peak in the amplitude image as expected and also contain a slight asymmetry. The measured interaction length in the lateral 0° amplitude image was found to be 211 nm in congruent crystals and 181 nm in near-stoichiometric crystals, which is wider than the vertical signal widths. Although amplitude calibration in the lateral direction to a physical distance is not possible, the amplitude of the images in congruent or near-stoichiometric are always similar in magnitude. The lateral phase image contains too much noise to be of any use.

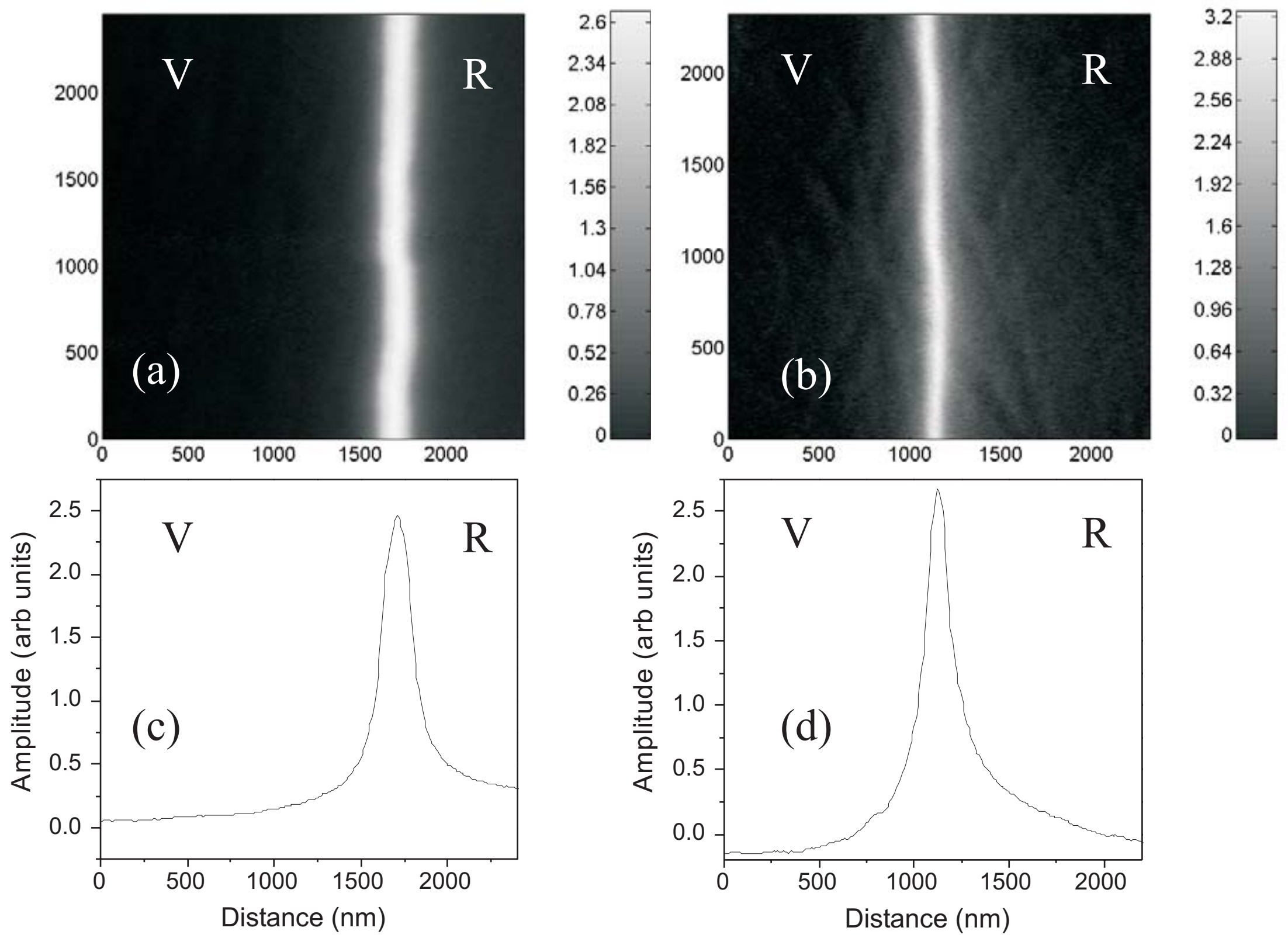

**Figure 7: Left-right PFM image (a),(b) and cross section (c),(d) for cantilever parallel to domain wall ($0^o$). Congruent lithium niobate (a),(c) and near-stoichiometric lithium niobate (b), (d).**

Shown in Figure 8 is the lateral image for the cantilever perpendicular to the domain wall (90° scan). This is a difficult image to obtain, mainly because the signal is small – about a tenth of the signal in the 0° scan – and because the measurement is very sensitive to the angle of the cantilever with respect to the domain wall. As the cantilever rotates from the perpendicular position to the wall, the signal amplitude begins to increase until the same shape and amplitude profile of the 0° scan is obtained at roughly 10° of rotation from the perpendicular position. The cross sectional curves are shown in Figure 8(c,d).

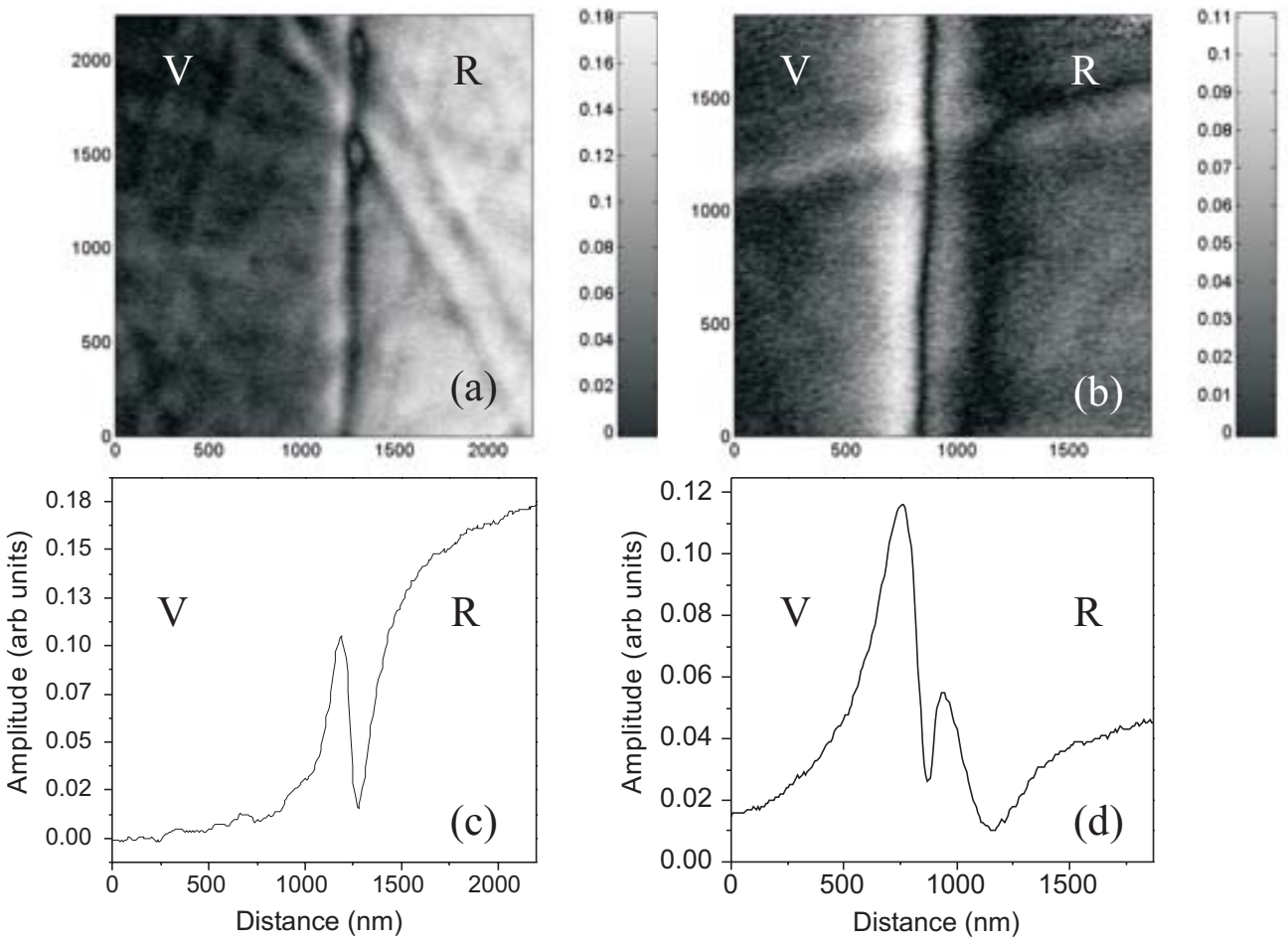

**Figure 8: Left right images in nm (a,b) and cross section (c,d) for cantilever perpendicular to domain wall (90$^o$). Congruent lithium niobate (a,c) and near-stoichiometric lithium niobate (b,d).**

The origin of the non-zero lateral 90º signals as evidenced in Figure 8(a) and (b) far from the domain wall is of unclear origin. One would expect the lateral signal to disappear far from the wall, as experimentally observed for the lateral 0º signals in Figure 8(a) and (b). The much weaker signal level of the 0º scans means the measurements much more susceptible to some of the inherent problems in using cantilever deflection scheme in which all degrees of motion of the cantilever are in some way coupled. This non-zero signal far from the wall appears to be a step like distribution in the signal superimposed on an anti-symmetric distribution present at the wall. The exact origin of the step-like contrast is unknown and requires further examination of the influence of other coupling effects. Considering the complex frequency response present for the vertical signal, investigation of this lateral signal might give clues as to its origin. However in this paper, we limit our discussion to the anti-symmetric distribution present at the wall and not on this step-like distribution.

This surprising local structure at the domain wall in the lateral 90° images in Figure 8(c,d) could have its origin in highly localized strains or distortions at the wall or be related to the defect dipoles. A further series of images was collected to show how local defect related fields could give rise to the contrast observed. As shown in Figure 9(a), a congruent crystal is poled from the virgin state (state 1) to a partially poled state (state 2), which is the state for most of the crystals imaged in this paper. In this situation, the reversed domains, R, contain defect dipoles with a less stable configuration than in the surrounding matrix virgin state, V. If we now partially reverse domains within the R state to a state $V_2$, we now have the original domain orientation similar to the virgin crystal, V, while the matrix domain state, R, has the unstable defect configuration. As shown in the schematic of Figure 9, this process creates domain walls separating domain states V and R, and well as walls separating states R and $V_2$ ($\equiv$V). As shown in Figure 9(b) and (c), the features observed in anti-symmetric behavior of the 90° lateral scans reverse their contrast in going from V-to-R versus going from R-to-$V_2$, clearly suggesting that these features arise from the presence of the frustrated defect dipoles. As mentioned before, a step-like signal is present in these images which appear to be larger in Figure 9(a) than (b). The origin of the step height is unclear, but it cannot be explained with the defect model and is more likely related to inherent cross coupling of the cantilever motion to another type of cantilever motion.

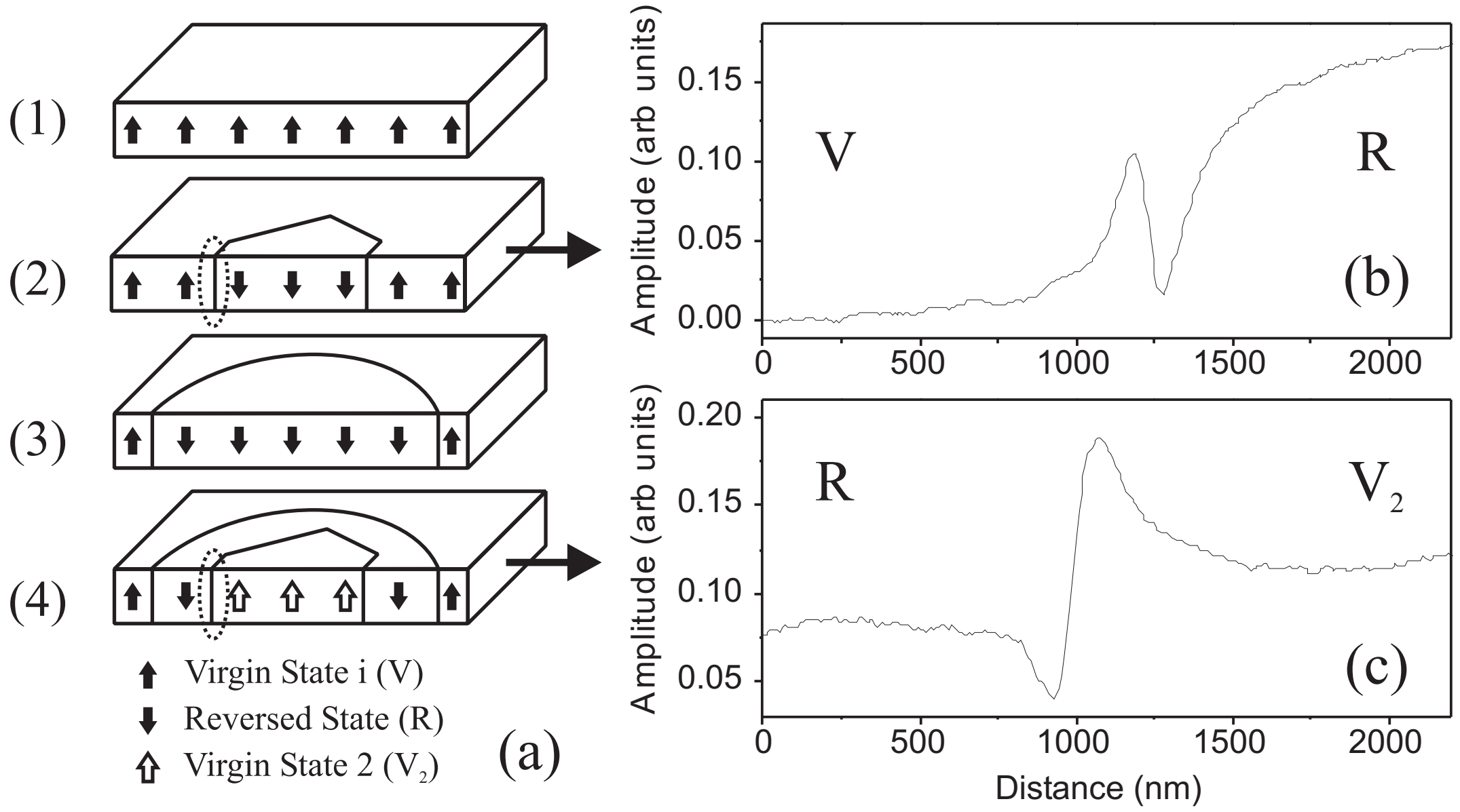

**Figure 9: Left right images for cantilever perpendicular to domain wall (90°) for two poling cases in congruent lithium niobate. (a) (1)-(4) shows the sequence of domain reversal in sample with (1) virgin state, (2) partial forward poling, (3) full forward poling under electrode, and (4) partial reversal where virgin state 2 is the same as the virgin state V with the addition of a poling cycle history. The domain walls circled in step (2) and (4) are imaged in (b) and (c) respectively.**

In summary, the differences between the near-stoichiometric and congruent piezoelectric responses at the domain walls support the premise that frustrated defects in the reversed domain (R) state affect the local electromechanical properties across a wall. The substantial reduction in the measured interaction widths between near-stoichiometric and congruent crystals indicates that the frustrated defects interact with the domain wall. The asymmetry always tails into the domain-reversed (R) area. Asymmetries in the vertical signal in congruent crystals are reduced with annealing and disappear in near-stoichiometric crystals where frustrated defects exist. In the lateral images, the differences between congruent and near-stoichiometric crystals are pronounced and of presently unclear origin. Next, we attempt to understand these PFM images more quantitatively using modeling.

**E. Electrostatic State of Surface**

Contribution to the domain wall contrast can also arise from the electrostatic state of the crystal surface, indicating perhaps a gradient in the charge compensation mechanism across the domain wall. Initial experiments were preformed using complementary noncontact techniques of Electric Field Microscopy (EFM) and Scanning Surface Potential Microscopy (SSPM) which probe the electrostatic state of the sample surface.[38, 39] In EFM the changes in the cantilever oscillating frequency caused by the force gradient above the surface are measured. EFM has been used successfully to determine the sign and density of surface charges in bulk TGS,[40, 41] GASH, [42, 43] PZT[44], and $BaTiO_3$.[30] SSPM uses electrical bias on the tip to null the potential difference between the tip and surface and allows high (~mV) potential resolution that has been successfully used to image domain walls in $BaTiO_3$[45, 46] and KTP.[37] SSPM and EFM imaging were preformed on a Digital Instruments Dimension 3000 NS-III using metal coating cantilevers of various resonance frequencies from ~60 kHz up to ~315 kHz. The lift height for both imaging techniques were varied between 10-200 nm above the sample surface. EFM measurements were taken with a series of bias voltages from -12 to 12 volts and the SSPM images were taken with an oscillating voltage of 5 volts peak. The domain wall appeared as a faint dark band in the optical microscopy used to focus and position the AFM cantilever which allowed domain walls to be located. However, *no* measurable difference across the domain wall was observed. The system was calibrated using a silicon substrate with with chrome interdigital surface electrodes across which various voltages were applied and measured. It was observed that below 50mV, no EFM images of the electrode could be observed. We therefore conclude that the surface potential difference, *if any*, across a domain wall in lithium niobate is *less than* 50mV. The potential difference between two adjacent *c*+ and *c*- domains has been measured as 155 mV in $BaTiO_3$[30] and 40 mV in KTP.[37] Measurements of potential screening on $BaTiO_3$ and charged grain boundaries in $SrTiO_3$ indicate that the lateral resolution is limited to ~300 nm related to the non-contact nature of the measurements.[47, 48]

## IV: MODELING PIEZOELECTRIC RESPONSE IN PFM

### A. Electric Field Distribution at the Tip

One of the primary unknowns in understanding a PFM image is the distribution of the electric field under an AFM tip with a small radius of curvature that is in contact with a ferroelectric surface (0.1 to 1 nm separation) idealized in Figure 10.

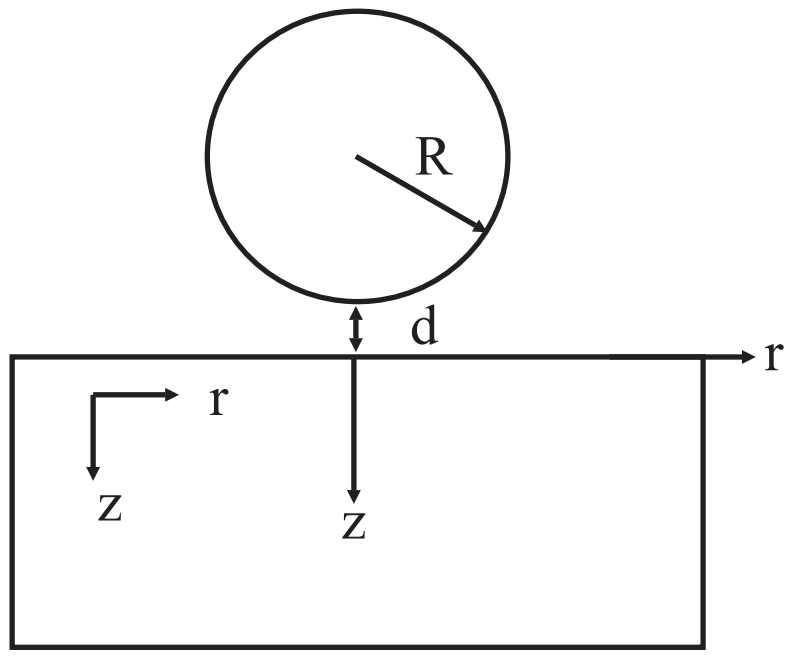

**Figure 10: Geometry of idealized AFM tip over anisotropic dielectric material.**

The approach taken in this paper is to use analytical solutions that describe an ideal electrostatic sphere-plane model. The limitations to this approach are discussed near the end of this section. However, these estimations are still useful and will be used as the input for the modeling in the next section. The first step is to determine the capacitance between a charged sphere and a dielectric material, given in der Zwan [49] as

$$C = R\sinh\alpha \sum_{n=0}^{\infty} \left( \frac{\sqrt{\varepsilon_z \varepsilon_r} - \varepsilon_1}{\sqrt{\varepsilon_z \varepsilon_r} + \varepsilon_1} \right)^{n} \frac{1}{\sinh{(n+1)\alpha}} \qquad \text{Equation 4}$$

and

$$\alpha = \cosh^{-1} \frac{R+d}{R} \qquad \text{Equation 5}$$

where units are CGS, and $R$ is the radius of the sphere, $d$ is the separation between the sphere and surface, $\varepsilon_r$ and $\varepsilon_z$ are the dielectric constants in the radial and $z$ direction respectively and $\varepsilon_l$ is the exterior dielectric constant (air in this case). Using the calculated value for capacitance the necessary charge, $Q$, for a given voltage can be found from $Q=CV$. The voltage and electric field distribution within the anisotropic dielectric sample can be found using the model given by Mele [50] as

$$E_z(r,z) = \frac{2Q}{\varepsilon_0(\sqrt{\varepsilon_r \varepsilon_z}+1)} \frac{1}{\gamma} \frac{z/\gamma - (R+d)}{\left(r^2 + z/\gamma - (R+d)^2\right)^{3/2}} \qquad \text{Equation 6}$$

$$V(r,z) = \frac{2Q}{\varepsilon_0(\sqrt{\varepsilon_r \varepsilon_z}+1)} \frac{1}{\sqrt{r^2 + (z/\gamma - (R+d))^2}} \qquad \text{Equation 7}$$

$$\gamma = \sqrt{\frac{\varepsilon_z}{\varepsilon_r}} \qquad \text{Equation 8}$$

where $E_z$ is the electric field in the z direction, $r$ is the distance coordinate parallel to the surface, $z$ is the distance into the sample, $Q$ is the calculated charge from the previous step, $\varepsilon_r$ and $\varepsilon_z$ are the dielectric constants of the anisotropic material in the radial and z direction respectively, $R$ is the radius of the AFM tip, and $d$ the separation between the sphere and surface.

Using values found in the literature and specifics for our tip geometry, with $\varepsilon_z$ = 28.1, $R$ = 50 nm, $d$ = 1 nm, the capacitance is calculated as $1.44 \times 10^{-17}$ F. With an imaging voltage of 5 volts, the resulting charge is $7.20 \times 10^{-17}$ C. From this value, the maximum electric field and voltage directly under the tip is $1.738 \times 10^{7}$ V/m and 0.51 V respectively. The distance into the sample where the field falls to $1/e^2$ value is 52 nm in the depth ($z$ direction), and 88 nm on the surface ($r$ direction). The overall normalized field ($E/E_o$, where $E_o$ is the maximum field) and voltage ($V/V_o$, where $V_o$ is the maximum field) in the sample is shown in Figure 11. These show the effect of field enhancement due to the small radius of curvature, as well as the quickly falling potential for even short distances from the tip. It is interesting to note that even a small imaging voltage of 5 volts results in a large electric field in the sample. The peak field generated in the sample using this model is only slightly below the coercive field of the congruent material ($2.2 \times 10^{7}$ V/m). If one considers a similar distribution in near-stoichiometric crystals, the coercive field ($4.0 \times 10^{6}$ V/m) is actually exceeded for a finite volume of crystal. This volume is an oblate spheroid with radius on the surface of 66 nm and penetrating into crystal a depth of 32 nm.

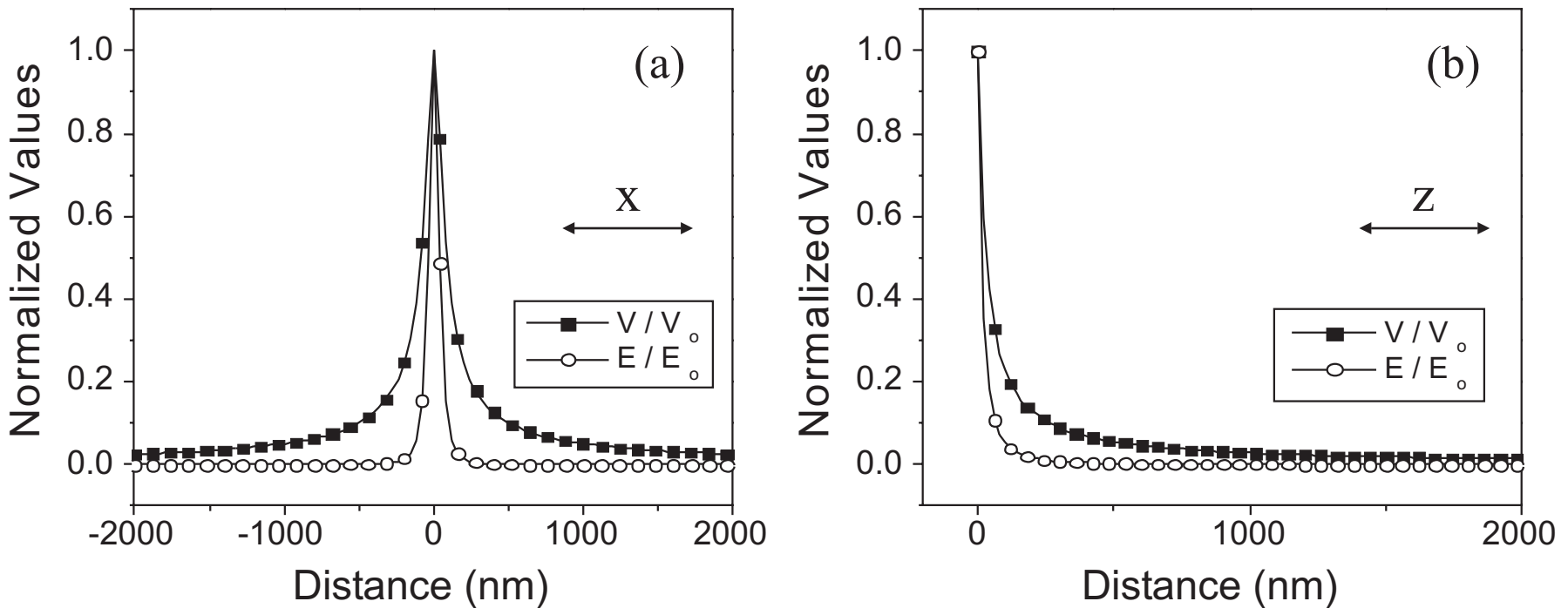

**Figure 11: Normalized voltage ($V/V_o$) and field distributions ($E/E_o$) on sample for imaging voltage of 5 volts separated 1 nm from dielectric surface where $V_o$=0.51 V and $E_o$=1.74 x $10^7$ V/m. Sample surface (a) and cross section (b).**

However, domain reversal is *not* occurring during the imaging process. When the maximum imaging DC voltage (5 V) is applied to the sample through the tip for periods of time up to 1 hour, no domain creation is observed. Similarly, Terabe have reported AFM tip poling of stoichiometric lithium niobate, and demonstrated that the process requires a time of at least one second to form stable domains for even a very high DC voltage (40 V) across a 5 μm thick crystal which generates a field 8 times higher than used in our imaging (1.4x$10^8$ V/m under the tip).[51] This switching time required is therefore much longer than the time for which the peak imaging voltage of 5 volts is applied to the sample (<25 μs). The coercive fields at such frequencies are unknown but trends show that coercive field increases with increasing frequency.[52]

There are several limitations to this distribution model. Recently, in several papers by Kalinin[29, 53], the imaging process in PFM can be separated into two distinct regions, the weak indentation limit, where the contact region between the sharply curved cantilever tip and the sample surface is a point contact, and strong indentation limit, where significant indentation of the sample surface by the tip increases the contact area and give rise to similar tip and surface potentials. Fields in the sample immediately under the tip in the strong indentation limit are most likely higher than in the plane-sphere model used here. However, the modeling in this paper is assumed to follow the weak indentation limit for FEM modeling simplicity. We feel this is justified considering the set point deflection of the PFM feedback loop for the images taken in the study were set to 0, and that the field distribution for model which includes indentation effects reduces to the point charge model for larger separations from the tip.[53] The inclusion of the electromechanical coupling effects would improve the modeling, but for the initial FEM modeling the sphere-plane model is a good first approximation.

In addition to the modeling uncertainty, the exact nature of the fields in the sample can be affected by surface and material properties. Issues include bound polarization charges, water, or other adsorbents on the surface, as well as a possible reconstructed ferroelectric surface layer with different properties than the bulk material (the so called "dead" layer).[9, 54] Since the exact nature of the surface is currently unknown, the proposed model here will be used as the maximum "ideal" field and will be used in the finite element modeling in following sections. The actual piezoelectric surface displacements calculated can then be treated as the "maximum" displacements that can be expected corresponding to these fields. The qualitative behavior of the piezoelectric responses across a wall can be predicted and compared with experiments.

**B. Finite Element Modeling**

Finite element modeling (FEM) of the sample surface under an electric field applied through an AFM tip was performed using the commercial software ANSYS [55].

Using a 10-node tetrahedral coupled field element with four degrees of freedom per node, the voltage and displacement in the *x*, *y*, and *z* directions in a slice of lithium niobate material under an applied electric field was simulated. The field distribution simulated a 50 nm tip separated 0.1 nm from the surface with a bias of was a ±5 volts, which is the same radius of curvature for the tips used in imaging. The material properties necessary for the simulation were the piezoelectric coefficients (18), elastic coefficients (21), unclamped dielectric constants (3), and the density, all found in the literature.[56] The physical dimensions of the simulated slice of material were 8 x 8 x 4 μm in the *x*, *y*, *z* directions respectively. The voltage distribution on the top and bottom surfaces was determined using the model described in the previous section and is shown in Figure 12, with a boundary condition on the bottom surface of zero net-displacement in the *z* direction. In each simulation, approximately a 13,000-element 19,000-node mesh was used in the solution, with the elements right below the applied voltage about 0.1 nm across. The model converged to the same solution when increasing the number of elements by 2 and 4 times. Although actual PFM experiments are performed with an alternating voltage (~35 kHz), only the static case was considered, i.e. the maximum displacements of the sample surface at peak imaging voltage (+/– 5 V). The FEM solution provides displacements of the sample surface at each node, $U_x$, $U_y$, and $U_z$. From these values, the distortion of the sample surface can be determined.

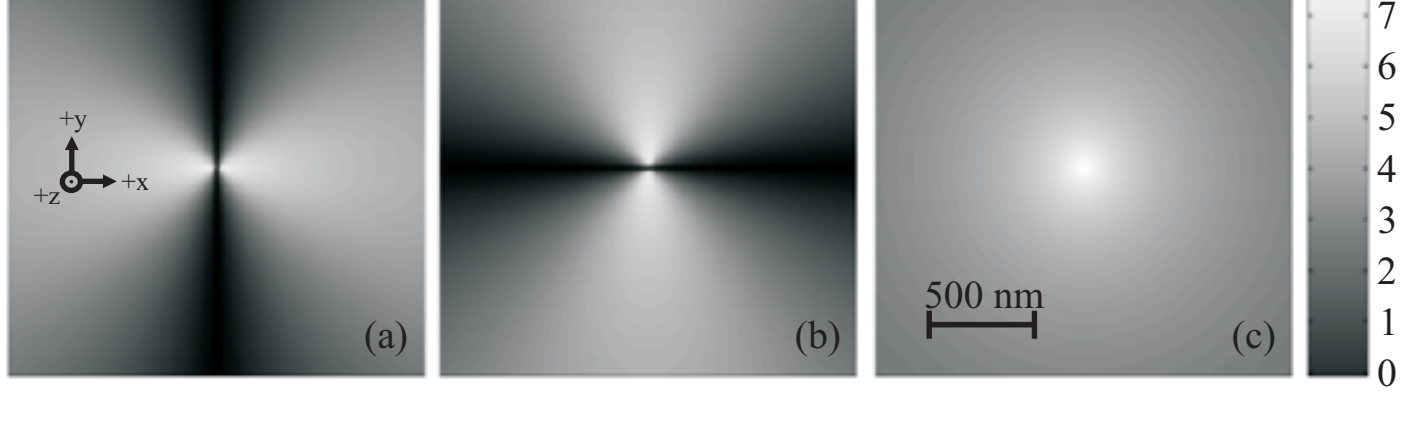

**Figure 12: Log10 of the electric field for the top surface of the lithium niobate used in finite element method modeling: *x*, *y*, and *z* components of electric field in (a), (b), and (c) respectively. Each plot is 2000 x 2000 nm.**

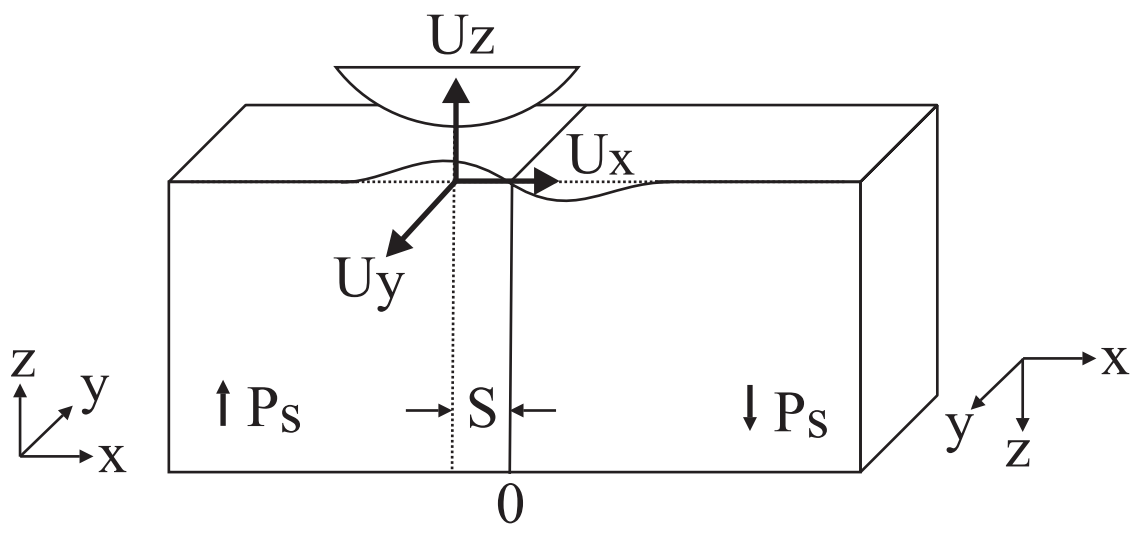

**Figure 13: Finite element modeling of the piezoelectric response across a domain wall in $LiNbO_3$. Probe is moved a distance, *S*, perpendicular to domain wall and displacement vectors describing surface displacements, $U_x$, $U_y$, and $U_z$, are determined.**

Two cases were modeled: the case of the field applied to (1) a uniform domain area on the surface of the sample and (2) a sample with the introduction of a single domain wall as shown in Figure 13. To model the domain wall, a solid block of material was divided into an *up* ($+P_s$) and a *down* ($-P_s$) domain by applying a coordinate system transformation to one half of the block as shown in Figure 13. The down-domain is obtained by rotating the crystallographic coordinate system of the up-domain by 180° about the *x*-axis, (2-fold rotation) thus resulting in $x \rightarrow x$, $y \rightarrow -y$, and $z \rightarrow -z$. The boundary between the two domains (at $x$=0) is a domain wall plane across which the properties change stepwise.

A series of simulations were performed as the fixed tip voltage was moved a distance, *S*, as shown in Figure 13, perpendicular to the domain wall for distances between –200 and 200 nm. Shown in Figure 14 are the surface distortions $U_x$, $U_y$, and $U_z$ calculated by FEM for 3 cases: (1) uniform domain with $S$=0, (2) a domain wall at $x$=0 and tip at $S$=0, and (3) domain wall at $x$=0 with tip at $S$=100 nm. Upon introduction of the domain wall in (d,e,f), the distortion on left and right sides of domain wall reverse compared to that in (a,b,c). The distortions become more complicated on moving the source away from the wall in (g,h,i). The tip was assumed to stay in the same position on the distorted surface, i.e. a tip at position ($x_1,y_1,z_1$) moves to ($x_1+U_{x1}$, $y+U_{y1}$, $z+U_{z1}$) where $U_{x1}$, $U_{y1}$, $U_{z1}$ are the distortion of the sample surface at the initial location of the tip.

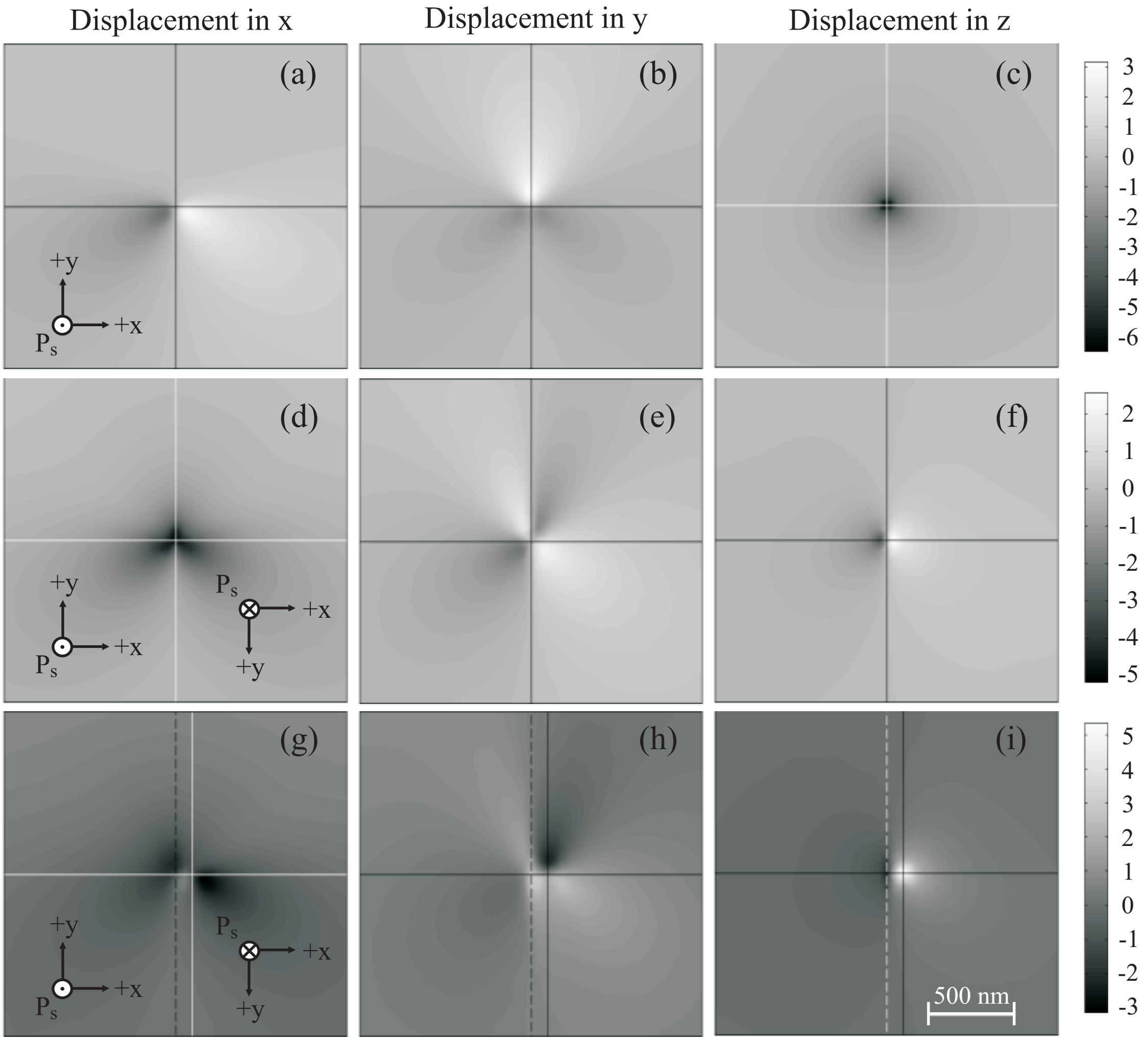

**Figure 14: Finite Element Method (FEM) calculations of surface displacements for +5 volts applied to the +$P_s$ surface for: a uniform domain with source at *S*=0 in (a,b,c), domain wall at *x*=0 and source at *S*=0 in (d,e,f), and domain wall at *x*=0 with source at *S*=100 in (g,h,i). Distortion $U_x$ is shown in column 1 (a,d,g), $U_y$ in column 2 (b,e,h), and $U_z$, in column 3 (c,f,i) with all distortions in picometers shown in common color bar on the right. Crosshairs indicate the position of tip, and the dotted vertical line indicates the domain wall. Each figure is 2000 x 2000 nm.**

### C. Simulation of Vertical Piezoelectric Signal and Experimental Comparison

To find the vertical piezoelectric signal from the FEM data the maximum expansion of the sample surface, $U_z$, underneath the tip was found for different positions, *S*, from the wall and is shown in Figure 15(a). This qualitatively mimics the PFM measurement as the distortion of the sample surface displaces the cantilever either up or down, and the lock-in amplifier measures this displacement. The amplitude signal measured in PFM is the peak-to-peak value of the sample displacement as shown Figure 15(b). It shows the expected result that away from the domain wall, surface expansion is the greatest and as the tip approaches the domain wall the magnitude of the oscillation goes through a minimum. The curves in Figure 15(a) were fit to curves of the form $A_o tanh(x/x_o)$ with a half wall width of $x_o$ = 58 nm. This curve was chosen and will be used for future curve fitting because it is identical in form to the change in polarization across the wall as given in Lines.[9] The minimum in the displacement at the domain wall is due to the mechanical interaction between the two oppositely distorting domains. The full-width-at-half-maximum of the FEM data gave an interaction width, $\omega_o$, of the domain wall as 64 nm. Far away from the wall, as simulated in the uniform domain case, the maximum amplitude of surface displacement was found to be 6.72 pm. The peak-to-peak amplitude oscillation value found by FEM is therefore 2 x 6.72 = 13.4 nm.

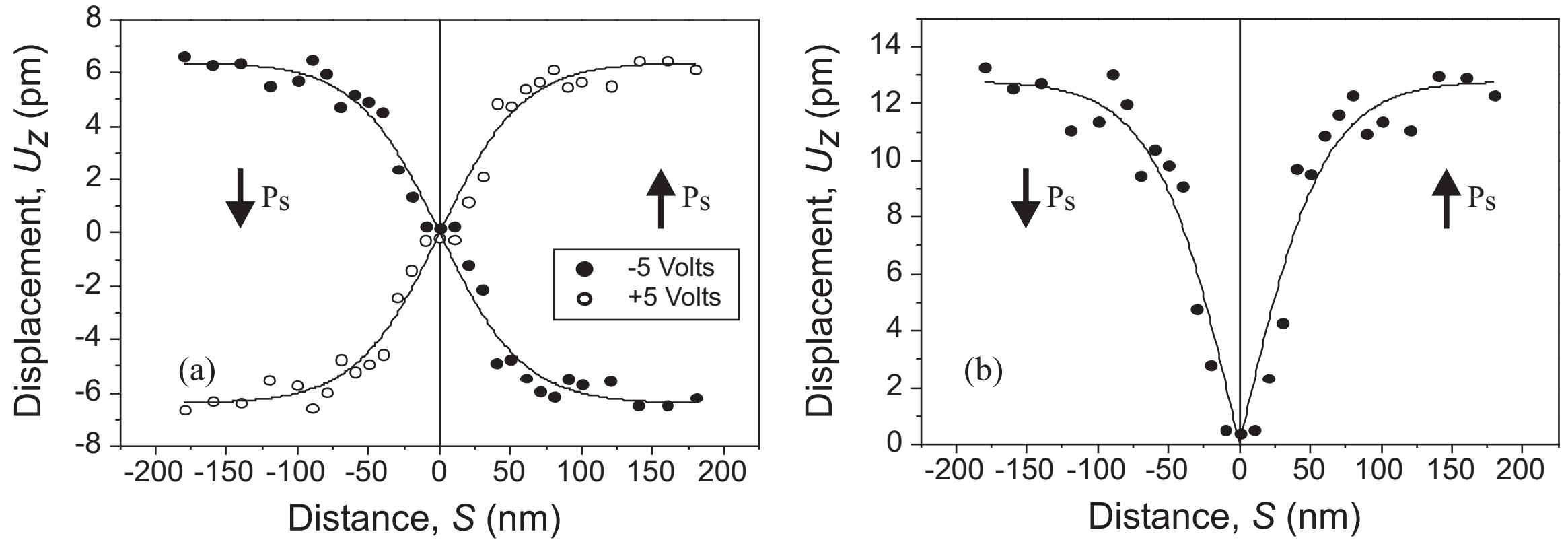

**Figure 15: Displacement $U_z$ underneath tip in FEM simulation as tip is moved across domain wall located at 0 nm. Each point represents the tip position relative to wall and maximum displacement of the surface. A best fit curve of the form $A_o tanh(x/x_o)$ is plotted as well. In (b) the absolute value of the difference between the two curves in (a) is plotted along with the absolute difference of the two best-fit curves in (a).**

The FEM modeling technique considers strictly the electromechanical behavior of the material. *It is important to note that the concept of nonstoichiometry is completely neglected in the FEM simulatio*n – all the simulation variables are the bulk material properties and the voltage distribution. Therefore, a comparison was made between the *near-stoichiometric* measurements and the FEM modeling.

Shown in Figure 16 is the measured vertical signal in near-stoichiometric LN along with the results from the FEM simulation. The magnitude of oscillation in simulation and measurement are very similar. The amplitude of oscillation experimentally measured away from the domain wall on both congruent and near-stoichiometric LN measure between 20-30 pm. This is of similar order-of-magnitude as the maximum oscillation predicted by simulation ~13.4 pm. If the separation is decreased to 0.1 nm (which increases the field in the sample) the static surface expansion would increase to 9 pm giving an oscillation of ~18 pm. The similarities of these results indicate that the electric field model and the finite element simulations give reasonable order-of-magnitude predictions. The forms of both curves in Figure 16 are similar, showing a dip in the signals at the domain wall returning to equal amplitudes on either side. However, the interaction widths, defined here as full width half maximum, are very different. The FEM simulation gives an interaction width, $\omega_o$, of 64 nm compared to the experimental $\omega_o$ of 113 nm.

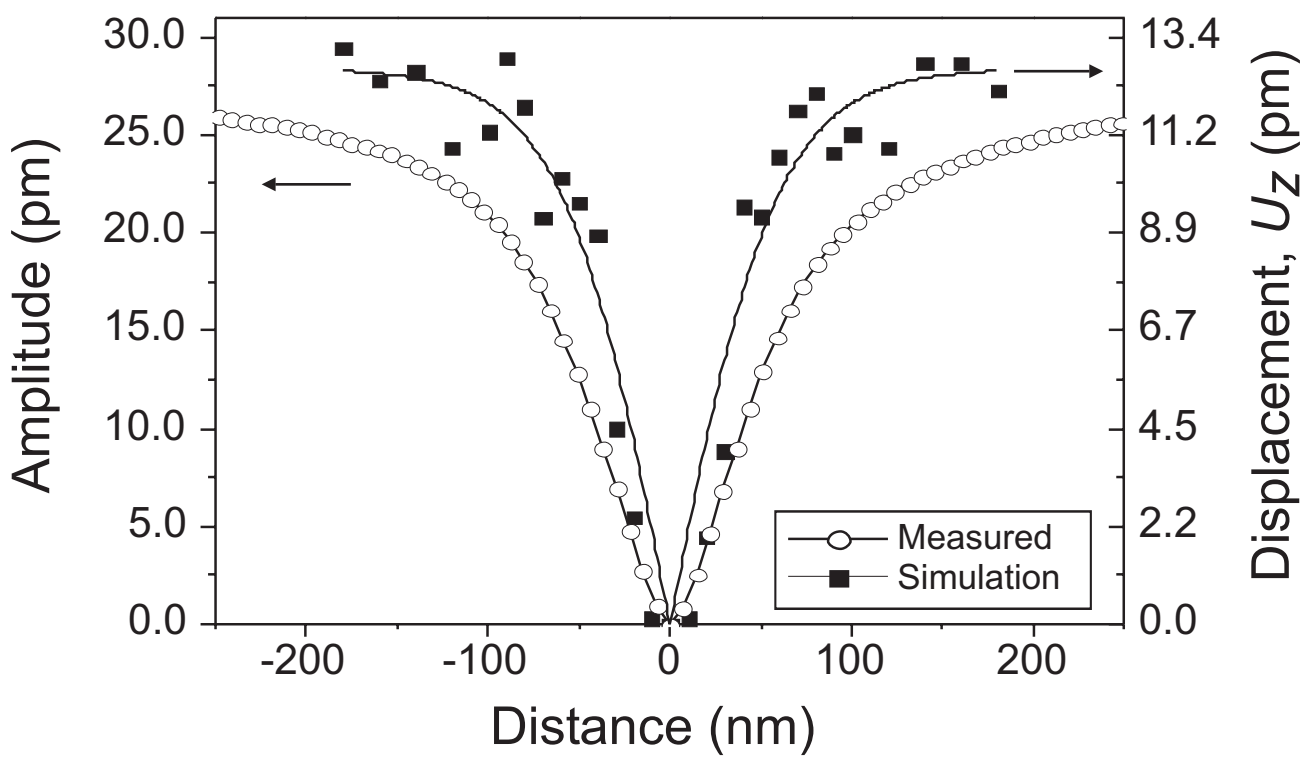

**Figure 16: Vertical amplitude signal on near-stoichiometric LN along with FEM simulation results with domain wall located at 0 nm. The simulation width is 65 nm compared to the experimental width of 113 nm.**

**D. Simulation of the Lateral Piezoelectric Signal and Experimental Comparison.**

The lateral signal for the cantilever parallel to the domain wall (*0° lateral scan*), shown in Figure 6(b) and (d), measures the torsion of the cantilever in the *x-z* plane, given by the slope of the sample surface in the *x-z* plane under the tip is shown in Figure 17(a). The results for a variety of tip positions, *S*, are shown in Figure 17(b). As the tip is moved toward the domain wall, the surface under the tip ceases to be flat, and starts to tilt as one side expands up and the other side expands down as pictured in Figure 17(a). When the voltage reverses polarity, the slope tilts the other way. In this way, a maximum in the lateral signal is measured in the domain wall area. The FEM data fit to the form $A_o tanh(x/x_o)$ gave the interaction width, $\omega_o$, as 59 nm, which is very close to the interaction width found from the *z* displacement analysis above (64 nm).

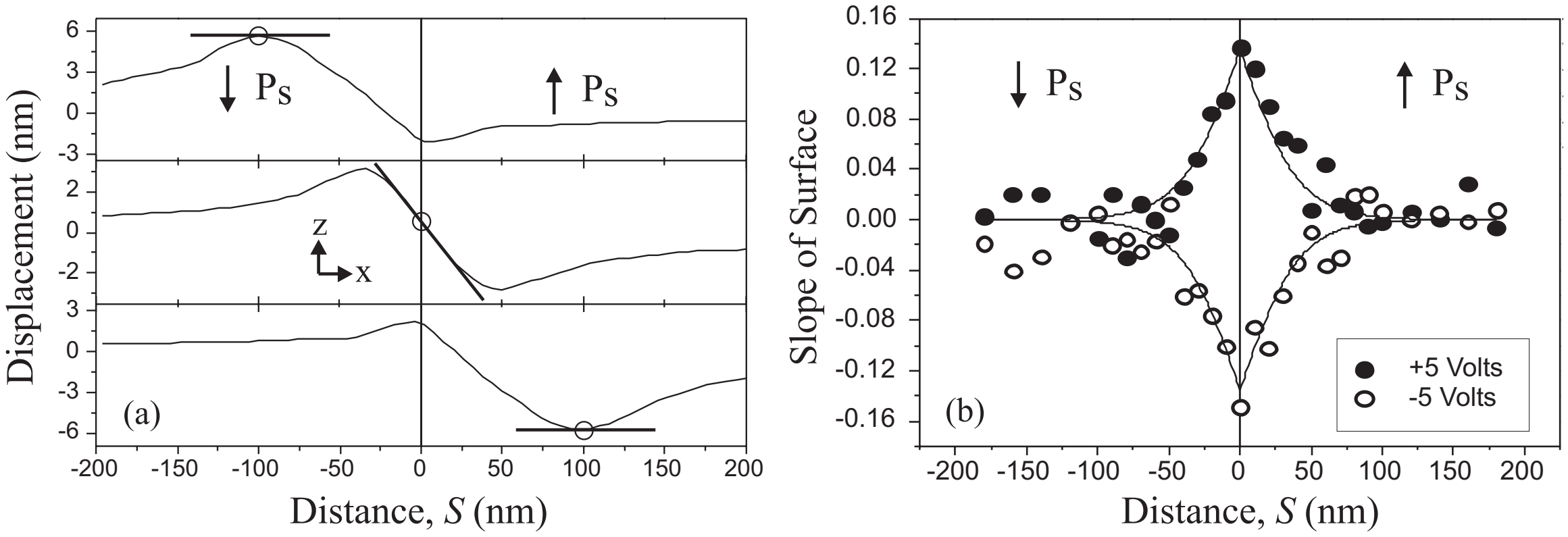

**Figure 17: FEM simulation of the lateral image amplitude with cantilever parallel to domain wall located at 0 nm (0$^o$ lateral scan). Shown in (a) are surface cross sections for –5V applied at 3 different tip positions (*S* = -100, 0, 100) and the slope of the surface at the tip position indicated by a circle. Shown in (b) is the slope of the surface under the tip for different tip positions, *S*, from the domain wall with a fit function of $A_o$ *tanh(x/x*$_o$*)*.**

A comparison of the 0° lateral scan results between simulation and experiments is shown in Figure 18. Although the forms of the curves are similar, (both showing a peak in signal at the domain wall), the FEM model predicts the interaction width, $\omega_o$, to be 45 nm compared to ~180 nm for the measurement. Since the experimental lateral signal cannot be calibrated, quantitative comparisons in the amplitudes cannot be made between simulation and measurement.

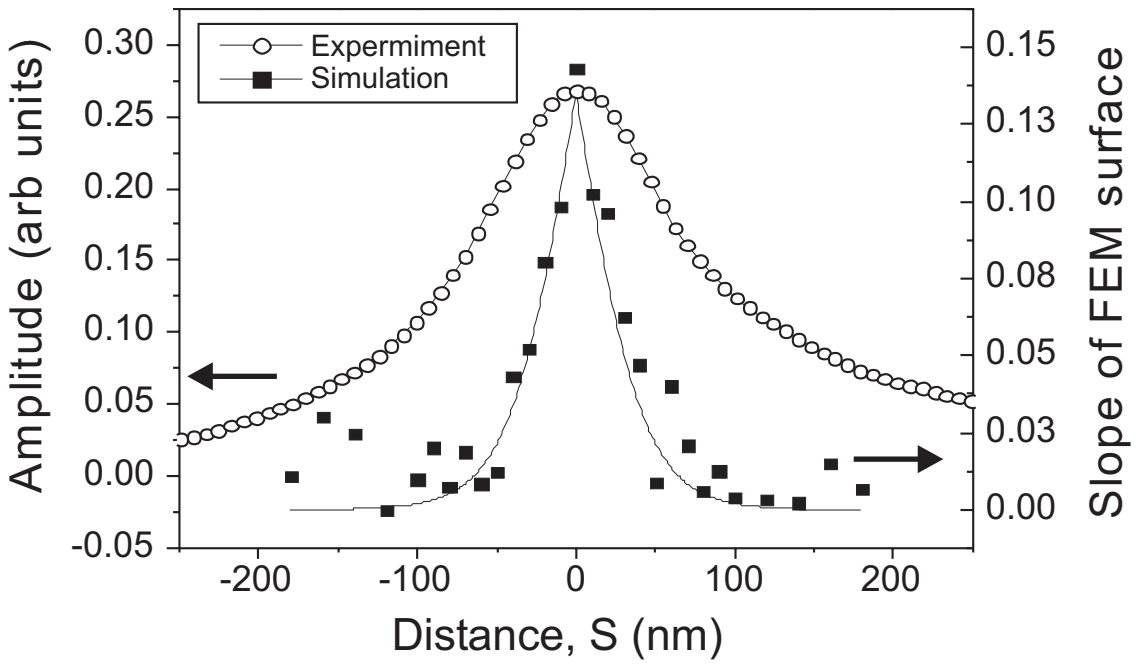

**Figure 18: Lateral amplitude signal for tip parallel to domain wall on near-stoichiometric lithium niobate along with FEM simulation results. The fit to the simulation data is the difference of the curves in Figure 17(b).**

The lateral signal for the cantilever perpendicular to the domain wall, (*90° lateral scan*) as shown in Figure 6(c) and (e), measures the torsion of the cantilever in the *y-z* plane. The *y*-axis switches orientation by 180° on crossing the domain wall. This tends to inhibit distortion in the *y-z* plane at the domain wall itself. Shown in Figure 19 is the evolution of the surface distortion in the *y-z* plane as the tip position, *S*, is moved away from the domain wall. At the wall, shown in Figure 19(a), distortion is minimal, and the slopes of the surface are also small. On moving away from the domain wall, the surface begins to distort again, mainly due to the movement in the z direction. The distortion in the *y*-direction is a pinching motion towards the tip when the surface expands up and an expansion away from the tip when the surface contracts down. At *S*=30 nm, shown in Figure 19(b), the surface is beginning to expand, with concave and convex bulges under the tip. The slopes of the surfaces are still small. For distances *S*=40 and larger shown (Figure 19(c,d)), the concave and convex bulges disappear, and the surface expands fully. When the surface expands, the tip is on top of a peak and measures the maximum slope as the tip is strongly influenced by displacements in the *y* direction that cause torsioning of the tip. However, when at the bottom of the depression, displacements in the *y* direction have less of an effect because the tip in a trough, and cannot easily torsion. Even though the electric field is symmetric about the *x*- and *y*- axes, the resulting distortions are not symmetric about the *x* or *y*-axis due to the 3-fold symmetry. An example is shown in Figure 16(b) where the displacement in *y* has three lobes. Any slice along this surface along the *y* direction will yield slightly more displacement on the upper half of the slice than on the lower side. This gives a net "bulge" along the *y* direction shown in Figure 19(c,d). The tip then follows the slant of this bulge, which tilts the cantilever at the peak preferentially toward one side for a given bias, and opposite for the opposite bias.

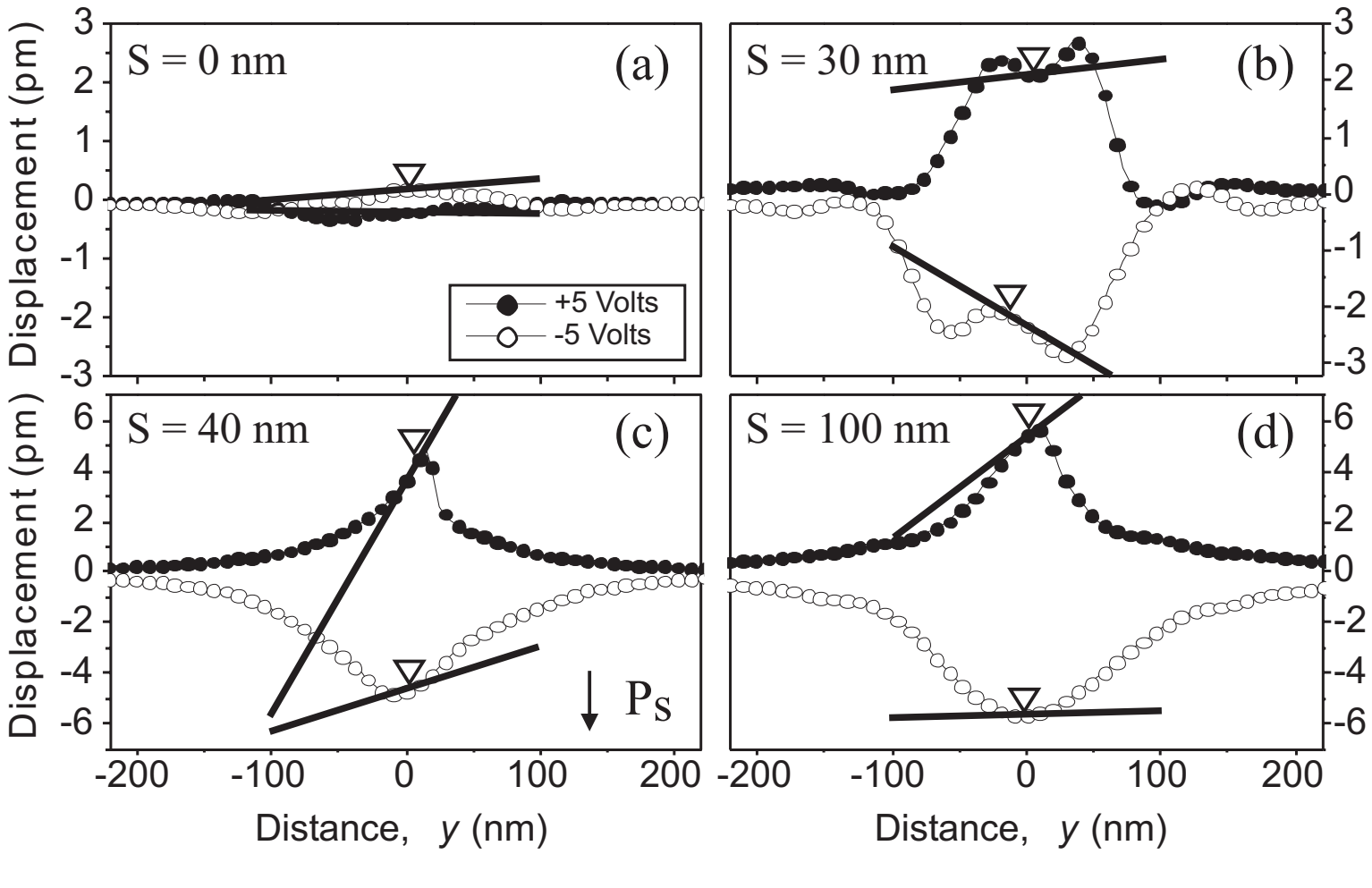

**Figure 19: Evolution of surfaces in *y-z* plane for different tip positions *S* from domain wall (at *x*=0) which is parallel to the plane of the plots. The slope of the curve at the position of the tip is the lateral signal imaged when the cantilever it**

**perpendicular to domain wall (90° lateral scan). The triangle represents tip position.**

The slopes of the surface in the *y-z* plane are plotted in Figure 20. They are very different between the positive and negative bias voltage, showing a peak above the down-domain for positive bias, and peak above the up-domain for negative bias. Since an oscillating bias is used in PFM imaging, the resulting signal from the expansion is shown in Figure 20(c) that shows a minimum at the wall and peaks slightly away from the wall.

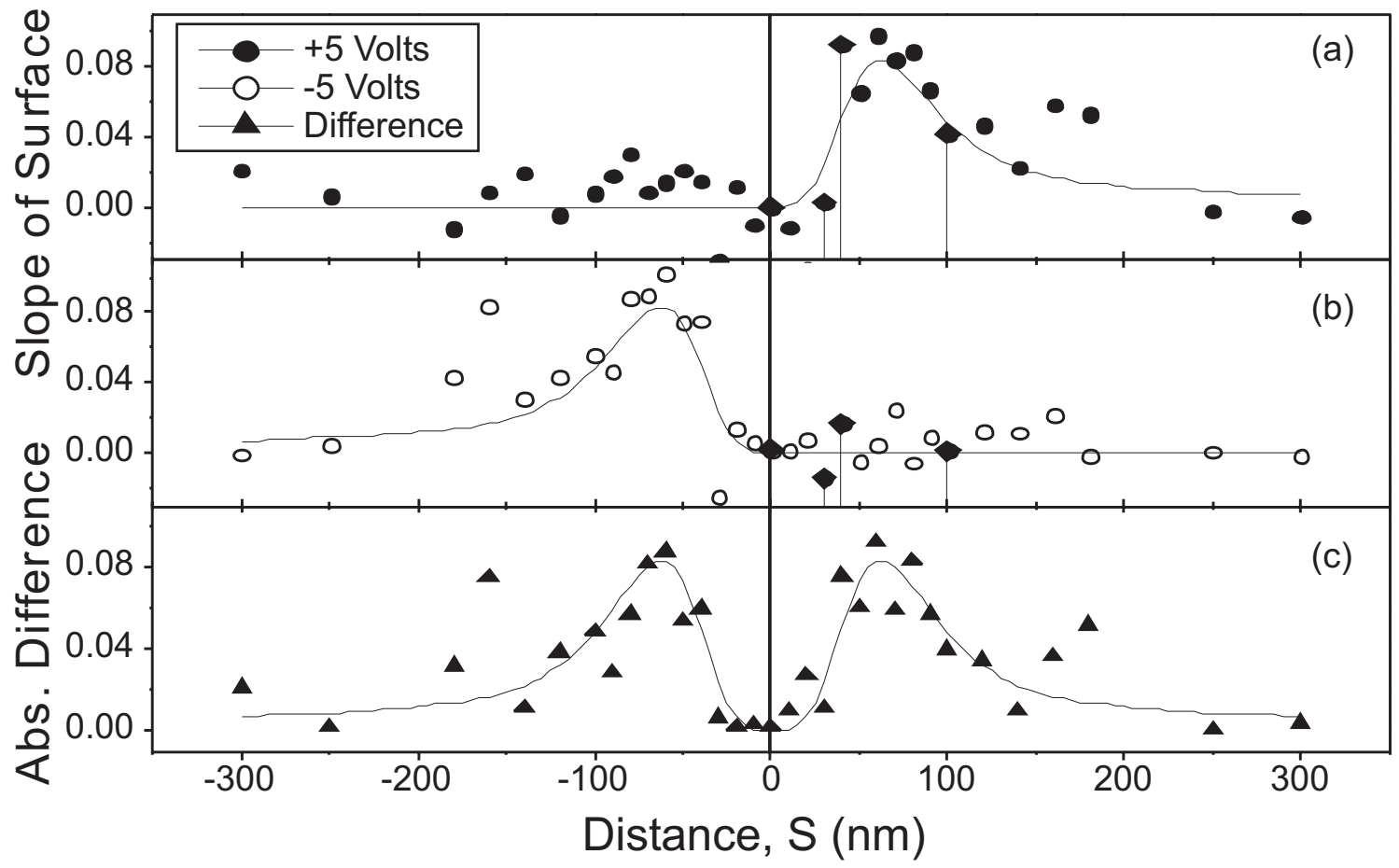

**Figure 20: FEM simulation of the lateral image with tip perpendicular to domain wall (90° lateral scan) located at *x*=0 nm for +5 V (a) and –5 V (b). Shown in diamonds with drop lines are the slopes to the surfaces shown in Figure 20. Shown in (c) is the magnitude of the difference between the two curves in (a) and (b) that is measured by experiment.**

The lateral 90° signal for simulation and measurement is shown in Figure 21. Although the FEM simulation data is noisy, a trend can be seen of a double peak with a minimum at the domain wall. This form is qualitatively similar to the data from the experimental measurement. The FEM simulation suggests that the signal measured in this direction is due to asymmetric bulging in the *y-z* plane that switches orientation on either side of the domain wall.

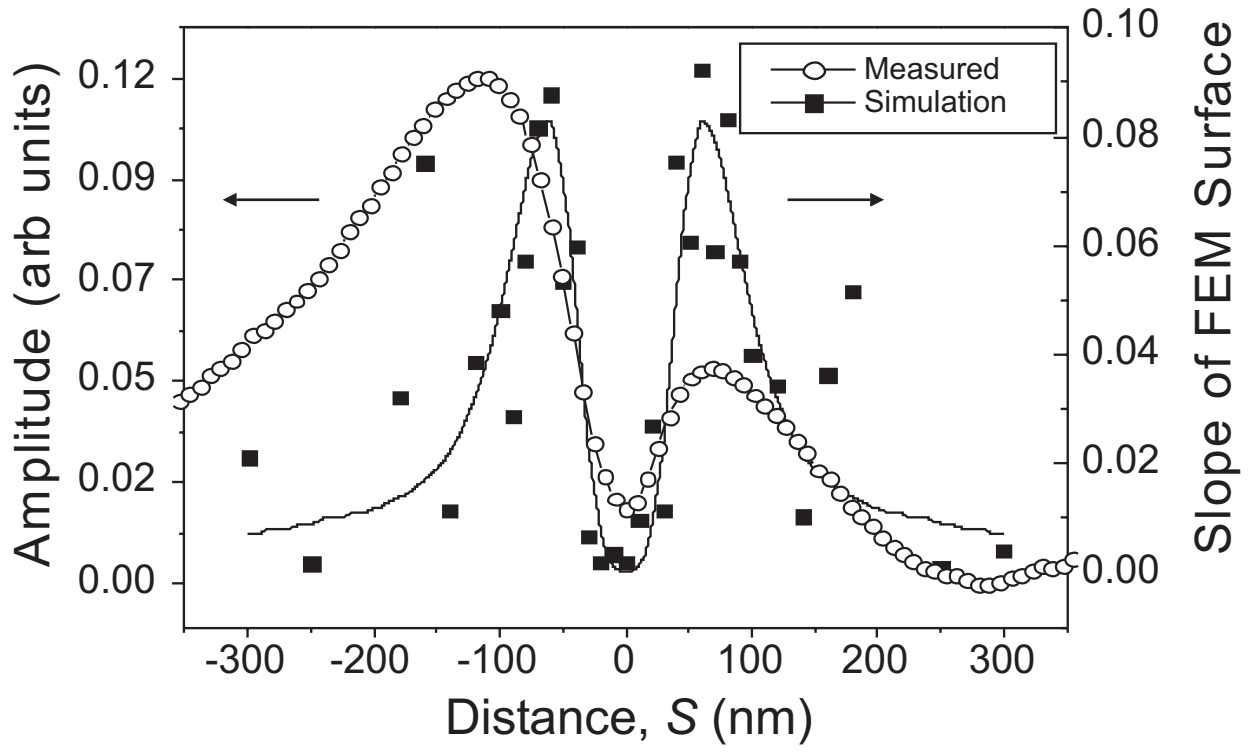

**Figure 21: Lateral image amplitude signal for tip perpendicular to domain wall (90$^o$-lateral scan) on near-stoichiometric LN along with FEM simulation results.**

It should be noted here that the acquisition of this profile is the most difficult for both the FEM modeling as well as the experimental measurement. The lateral FEM signal information, where the slope to the distorted surface was used, has larger associated errors than the vertical signal. Aside from the inherent numerical errors due to discretization of continuous functions into finite elements, the majority of error came from sampling. First, the slope of the distorted surfaces at the probe point is required, which is the tangent to the surface at one point. While the displacements themselves are continuous, the first derivative of the surface displacements are not necessarily continuous and are very sensitive to conditions around the point sampled. Several node points were considered in the determination of the tangent values. In the case of the lateral 90°-signal plane, the distortions are so small in the *y* direction that behavior is dominated by the z-signal. Amplifying the *y* displacement by a factor of ten allows a trend be seen in the data. In this way, the *y-z* signal data should only be used to illustrate a *possible* trend.

## V. DISCUSSION

In order to place the comparisons in proper context, there are several limitations and assumptions present in the finite element model that need to be discussed. The first is that the voltage and electric field in the sample surface are assumed to be identical to the analytical solution given in the previous section. This is an idealization of the physical reality, since the absolute field values at the surface depend on the surface structure and conditions that are not precisely known. In addition, it is assumed that the physical properties of the sample determined from a bulk crystal (i.e. piezoelectric coefficients) apply at very small length scales and are valid for describing small volumes near or on the surface. The actual imaging technique uses an oscillatory voltage that can introduce resonance effects into the measurements, whether in the cantilever, sample surface, or both. The static FEM simulations ignore these effects.

Despite these limitations, the FEM simulations can be used to determine two pieces of information: the magnitude of the sample oscillations and the interaction width of the wall. The quantitative surface displacements can be considered to be the maximum values that the surface can possibly expand. Their values are of the same order of magnitude as the measured displacements (13.4 pm for the FEM simulations compared to ~20-30 pm for the experiments).

The measured interaction width at a domain wall ($\omega_o$~113 nm) in the experimental PFM images is twice as large as the FEM model (~ 64nm). There are several factors that contribute to the interaction width of the wall. This width should be thought of as the upper limit of the interactions at the wall and include contributions from the applied field (magnitude and distribution), tip geometry (radius), surface effects (charge distribution), and sample properties (dielectric, piezoelectric, and elastic constants). Of these, the FEM simulation only models the electromechanical behavior of the sample; therefore the dip in amplitude at the domain wall in the simulation is due only to the electric field distribution, the strain compatibility, and the mechanical coupling of the two oppositely expanding domains. We next explore some of these contributions.

One limiting factor to the interaction width of the domain wall is the inherent mechanical coupling present between the oppositely expanding domains. The width of the transition from full expansion to full contraction depends on some combination of the elastic and electromechanical constants of the material and the thickness of the sample. An exact analytical solution to this problem can be approached using Ginzburg-Landau-Devonshire (GLD) theory;[57, 58] however, consideration of the sample surface and field distribution complicates this problem greatly.[59-61] To get a numerical solution from FEM, a *uniform electrode* was defined on both the top and the bottom surfaces of the finite element model discussed earlier, so that there was a uniform field distribution in the bulk of the material. This uniform field is a simulation of the limiting case where the tip radius $R \to \infty$. It was found from FEM modeling that for a uniform electric field, the inherent electromechanical width across a single 180° wall is independent of the applied electric field for a sample of constant thickness as shown in Figure 22(a). While the maximum surface displacement increases linearly with the field as expected, the electromechanical width remains the same for a given crystal thickness. Also, the electromechanical width is linearly related to the sample thickness as shown in Figure 22(b) for a fixed uniform electric field value. An "intrinsic" parameter can be defined as the dimensionless ratio $\omega_{pi}/t$, which is independent of external field or sample thickness. This parameter in $LiNbO_3$, which relates the electromechanical width to the sample thickness *(t)*, has a value of $\omega_{pi}/t$~0.16. (Note that $\omega_{pi}$ is the FWHM wall width). For the 300 µm thick crystals as used in this study, the intrinsic electromechanical width for a uniform field is extrapolated to ~49 µm. This value, while quite wide, is supported by X-ray synchrotron measurements taken of a single domain wall in $LiNbO_3$ which shows long range strains of ~50 µm on the sample surface.[62, 63] Interaction widths scaling with the sample thicknesses have been experimentally observed in PZT thin films imaged by PFM where larger interaction widths were measured for thicker films.[44, 64] The scaling factor calculated from these PZT thin film measurements is ~0.09. Both the simulations and the experimental observations point to an ultimate limit to the resolution that is related to the electromechanical response of the material and the sample thickness. The intrinsic parameter, $\omega_{pi}/t$ must definitely related to $d_{33}$, $d_{31}$, $\varepsilon_{33}$, and $\varepsilon_{31}$. However, GLD theory of spontaneous strain widths in single *infinite* domain wall in lithium niobate (with no surfaces and no external fields) indicates that it is related to all piezoelectric, dielectric and elastic constants of the material.[65] One could reasonable expect a similar situation in the homogeneous case described above that has the added complexity of surfaces and external fields.

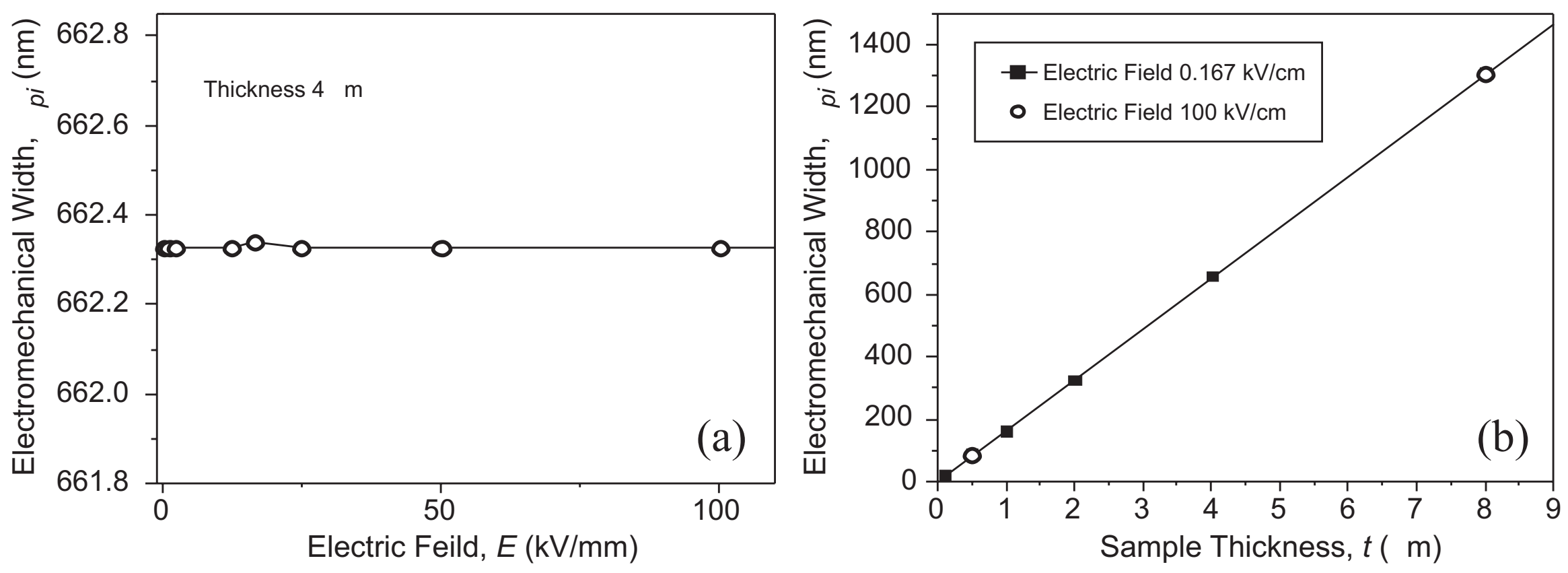

**Figure 22: FEM simulations of the electromechanical interaction width (FWHM), $\omega_{pi}$, under uniform electric field applied to samples for (a) varying electric field and constant thickness of 4 μm and (b) varying sample thickness and fixed electric field.**

By using a PFM tip electrode on one face, much higher PFM wall resolution is possible in thicker crystals due to the highly localized electric fields produced by the tip near the surface. To examine the influence of the tip radius and electric field effects in the sample, the FEM modeling was preformed for a variety of tip radii using the electric field model for a 5 V imaging voltage. The results of these simulations are shown in Figure 23(a). For tips larger than the 50 nm radius used in this study, the interaction width predicted by the FEM model is relatively insensitive to the radius. As the radius gets smaller than 50 nm, there is a sharp reduction in the measured interaction width.

In an attempt to understand Figure 23(a), the electric field distribution was found for a variety of tip radii. The crystal depth, *d*, below the tip, below which the electric field in the sample did not give rise to measurable displacement, was experimentally determined by finding the minimum applied voltage (0.6 V peak) that generated a signal the lock-in amplifier could measure. Using this voltage value, the peak field under the tip calculated from the analytical model is $2.9 \times 10^6$ V/m which is used as the field value for determining the electric field distribution of the oblate spheroid with a radius, *r*, on the surface and penetrating into crystal a depth, *d*, into the surface with a total volume, *V*. These values are normalized to the maximum values for each curve and are shown in Figure 23(b). The trends show the expected results that the peak electric field is enhanced for smaller radius tips and the distribution becomes more diffuse for increasing tip radii.

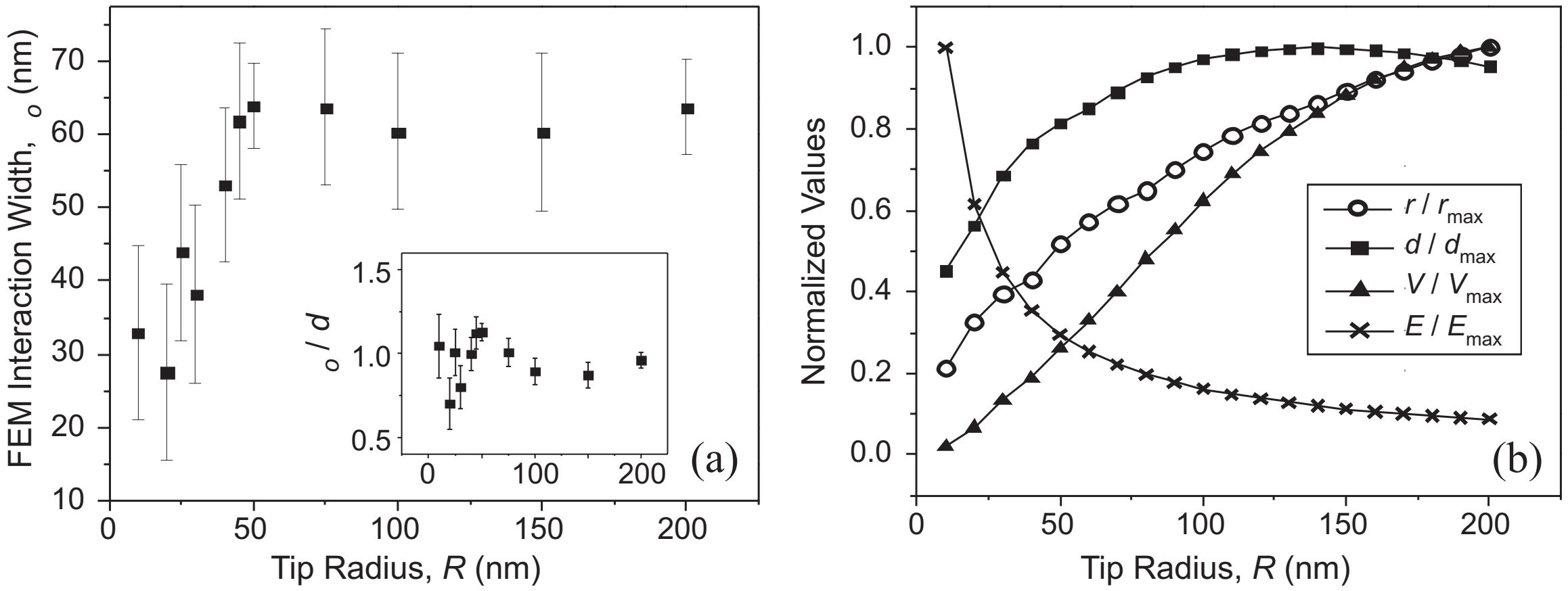

**Figure 23: (a) FEM simulations of interaction width, $\omega_o$, for a variety of tip radii, *R*. (b) normalized values of the maximum electric field under the tip and the field distribution for varying tip radii where the field falls to the experimentally determined value of 2.9x10$^6$ V/m below which no displacement could be measured. For normalization, $E_{max}$ = 5.88x10$^7$ V/m, $d_{max}$ = 69.6 nm, $R_{max}$ = 183 nm, and $V_{max}$ = 2.81x10$^{-21}$ nm$^3$ are used. Inset of (a) gives the engineering parameter, $\omega_o/d$, where *d* is given in (b).**

From Figure 23(b) the sharp drop off in the FEM calculated interaction widths do not correlate exactly with any of the calculated field distributions. The maximum field, *E*, under the tip is enhanced for smaller tip radii, *R*; however, this is unlikely to contribute to increased wall resolution because it is the distribution of the field that is important.

The flat region of Figure 23(a) roughly correlates with the depth data in Figure 23(b), which is in the range of $R$=60-200 nm with a mean value of $\omega_o$ ~70 nm. In an analogy to the $\omega_{pi}/t$ as defined before for the uniform electrode case, we can define an "engineering" parameter, $\omega_{pi}/d$ that is very approximately independent of the tip radius (within ± 26%). The parameter is only approximate, with the variation of the signal attributed to the nonlinear dependence of the electromechanical width and the penetration depth of the electric field on the tip radius.

In conclusion to the wall width issues, we can state that there exists a thickness dependent intrinsic electromechanical width to an antiparallel domain wall under uniform electrodes. This width can be substantially modified by choosing non-uniform fields using PFM tip geometry. Therefore, in a PFM measurement of antiparallel domain walls, which of these effects dominates depends, in general, on the tip geometry and the sample dimensions.

The possibilities of the larger interaction width in the PFM experiments as compared to the modeling could be related to surface effects not accounted for in the FEM modeling. If there were a "dead" layer on the surface that is paraelectric, caused by surface reconstruction or diminishing spontaneous polarization near the surface, this would introduce a distance between the voltage source and the piezoelectric material that would act to decrease the electric field in the piezoelectric portion of the sample. Similarly, the presence of a thin film of water on the sample surface would cause a broadening of the electric field. This was observed by Avouris in the oxidation of silicon surfaces with an AFM tip where it was necessary to replace the tip radius with much wider meniscus of water to model their results.[66] Both of these situations then broaden the electric field distribution. While the FEM model predicts relative insensitivity to broader electric field values, this none-the-less could be the origin of the broadening in the actual measurement.

Other possibilities for the domain broadening and asymmetry could be the electrostatic distribution on the surface around a domain wall. If we assume that compensation of the ferroelectric polarization is at least partially accomplished by surface charges adsorbed from the environment then there is a charged double layer on the surface. There would then be a diminishing or enhancement of the amplitude due to electrostatic interaction of the tip by the charged surface. The sign of this charge changes across a domain wall and would introduce a gradient in the electrostatic signal that would be present in the interaction width. We will examine two simple cases of screening – an under-screened surface, meaning net bound polarization charge remains on the surface, or over-screened surface, meaning net bound polarization is over-screened by the surface layer, as shown in Figure 3(c) and (d) respectively.

Let us examine a simple model of a domain wall at $x=0$ being scanned by a positively charged tip and only consider the spatial distribution the piezoelectric and electrostatic amplitudes, $A_{pi}(x)$ and $A_{es}(x)$, respectively. The variation of the signal on crossing a domain is given as a hyperbolic tangent which was used to fit the simulated vertical data in Figure 16(a). The total amplitude signal, $A_o(x) = A_{pi}(x) + A_{es}(x)$, as a function of distance, $x$, is then given as

$$A_o \tanh(x / x_o) = A_{pi} \tanh(x / x_{pi}) + A_{es} \tanh(x / x_{es})\, e^{i\theta} \quad \text{Equation 9}$$

where $\theta$ gives the phase relation between the electrostatic amplitude and the positively charged tip, and $x_o$, $x_{pi}$, and $x_{es}$ are domain wall half width widths. These are related to the interaction widths (FWHM) by $x_o = 0.91\omega_o$, $x_{pi}=0.91\omega_{pi}$, and $x_{es} = 0.91\omega_{es}$. There are two different types of electrostatic signals, $A_{es}$: one from an over-screened surface, $A_{es} = A_{ov}$, and one from an under-screened surface, $A_{es} = A_{un}$. The phase, $\theta$, is π for an under-screened surface and 0 for an over-screened surface. The variation across the wall for the piezoelectric and electrostatic signals is shown in Figure 24(a) where $A_{pi} > A_{es}$.

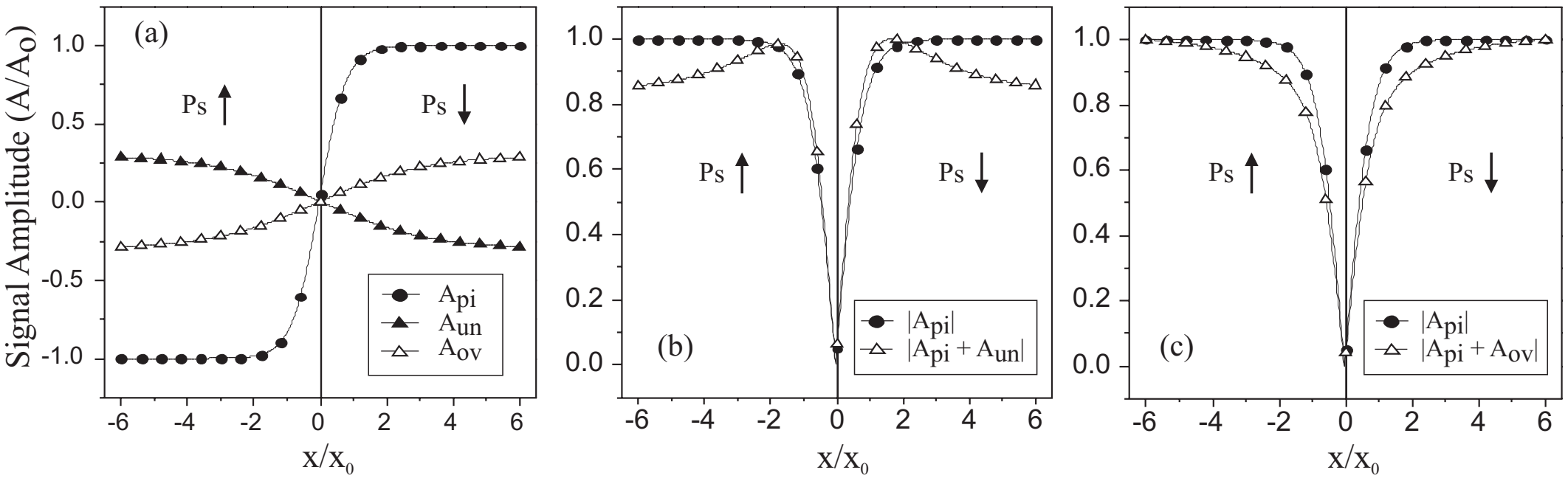

**Figure 24: Influence of static electrostatic gradient on the imaging of the vertical signal. (a) spatial distribution of amplitude and phase for a positive tip voltage for the piezoelectric signal, $A_{pi}(x)$, and electrostatic signal for an over-screened, $A_{ov}(x)$, and under-screened surface, $A_{un}(x)$. (b) Magnitude of the normalized amplitudes of the piezoelectric and the net piezoelectric and electrostatic signal for an under-screened surface and (c) Magnitude of the normalized amplitudes of the piezoelectric and the net piezoelectric and electrostatic signal for a over-screened surface.**

If the domain regions are under-screened, the electrostatic signal, $A_{un}$, will be contrary to the piezoelectric signal ($\theta$=π). Summing the two signals for an under-screened surface gives the net amplitude (the absolute value of equation 9) as shown in Figure 24(b). The net amplitude acquires a ridge around the wall, caused by adding the contrary signals. This ridge structure is not experimentally observed.

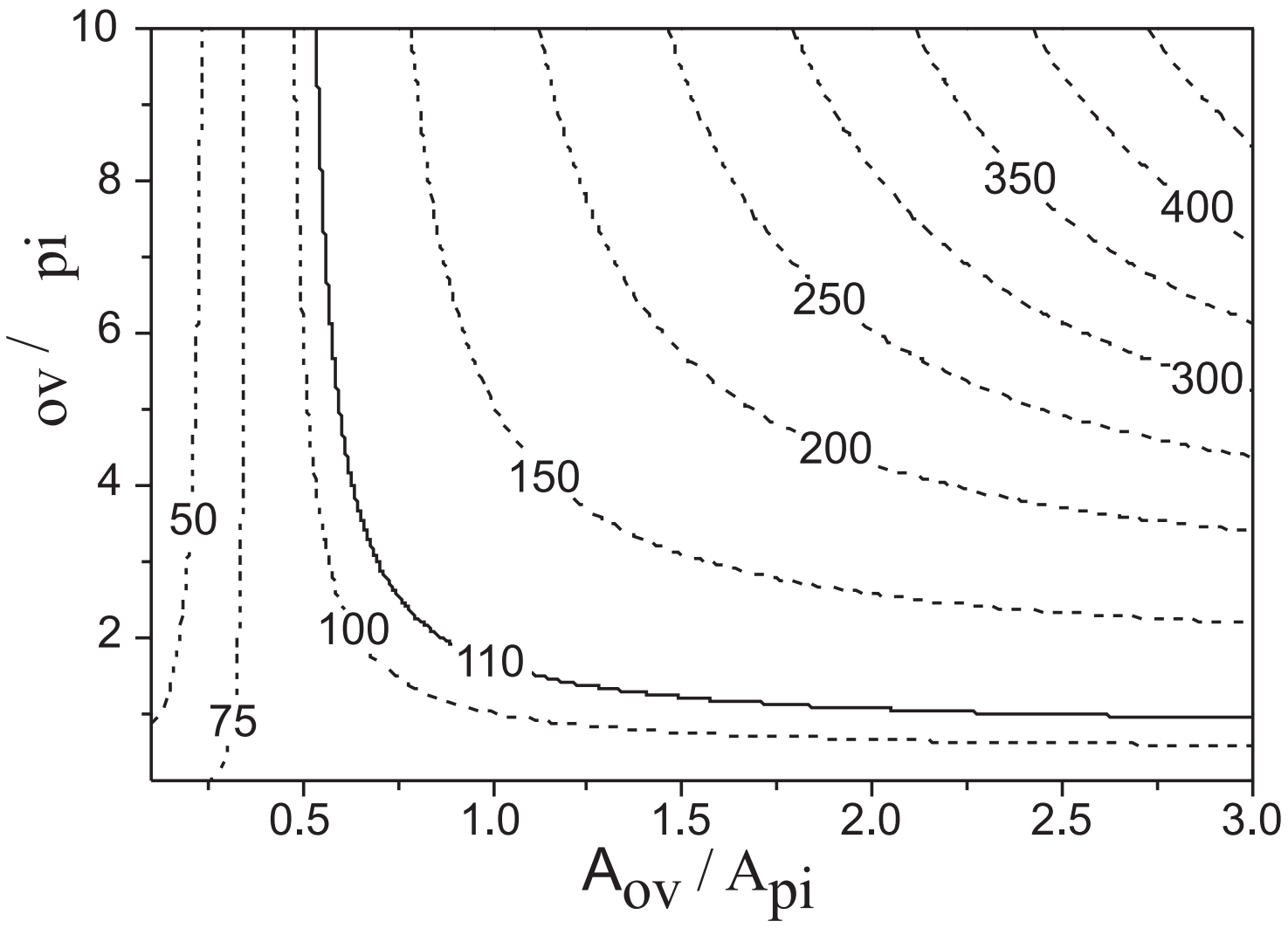

**Figure 25: Contours in nanometers of the full-width-at-half-maximum width for the combined piezoelectric and over-screened electrostatic signals versus the ratios of the electrostatic to the piezoelectric amplitude ($A_{ov}/A_{pi}$) and transition widths ($\omega_{ov}/\omega_{pi}$). The dark line indicates the experimentally measured interaction width ($\omega_o$~110 nm) on stoichiometric lithium niobate.**

If we let the surface be over-screened, the phase difference θ=0° in Eq. 9, and the resultant amplitude is shown in Figure 24(c). One can notice that the combined signal is wider than just the piezoelectric signal alone. The amplitude and transition width of the over-screened electrostatic signal can cause broadening of the net signal observed in PFM measurements. Plotting ratios of the amplitudes of the signals ($A_{ov}/A_{pi}$) and to the interaction widths ($\omega_{ov}/\omega_{pi}$) gives different values of the interaction width as shown in Figure 25. A variety of ratios can give a net interaction width equal to the experimentally measured width (~110 nm) assuming the piezoelectric interaction width is given by the finite element method simulation result (~65nm). For example, if $A_{ov}$ is equal to $A_{pi}$, then the $\omega_{ov}$ is approximately twice as wide as $\omega_{pi}$. Although over-screening can explain signal broadening, the mechanism for an overscreened surface is presently unclear. An over screened surface has been observed on reduced SAW grade $LiNbO_3$,[67] although comparison to congruent optical grade wafers is not easy due to the severely modified electrical nature of the reduced samples.

An estimation of the ratios of the signals can be made using the maximum possible value of the electrostatic surface potential difference (if any exists at all) across a domain wall of 50 mV, estimated from the SSPM and EFM measurements. Following the formulation of Hong,[27] the amplitude of the electrostatic signal, $A_{es}$, is approximately given by

$$A_{es} = -\frac{1}{k}\frac{dC}{dz}V_c V_{ac} \qquad \text{Equation 10}$$

where $V_{ac}$ is the applied oscillating imaging voltage, $k$ is the cantilever spring constant, $dC/dz$ is the capacitance between the tip-cantilever system and the sample surface, $V_c$ is the surface potential measured using SSPM. Using the electric field model in Section IVa**,** the $dC/dz$ term can be numerically calculated as $-1.73\times10^{-9}$ F/m at 0.1 nm tip separation. If $V_c$ is set equal to the *upper limit* of 50 mV estimated using EFM and $k$ is 12 N/m, the *upper limit* value of $A_{es}/V_{ac}$ is ~3.60 pm/V. The ratio $A_{pi}/V_{ac}$ calculated from the FEM modeling is given as 13.4 pm / 5 V which is ~2.7 pm/V. From the data collected in using EFM and SSPM, the nature of the surface screening cannot be determined. However, if we assume over screening and use the model above, the ratio $A_{ov}/A_{pi}$ is 1.34 which gives a ratio $\omega_{ov}/\omega_{pi}$ of ~1.85 from Figure 25. This gives an estimation of the electrostatic signal width as $\omega_{ov}$ ~120 nm (1.85 x 65 nm). If we take the upper limit of the piezoelectric signal, $A_{pi}$, as equal to the piezoelectric coefficient ($A_{pi}/V_{ac} = d_{33} = 6$ pm/V), $A_{ov}/A_{pi}$ is 0.6 which gives a ratio $\omega_{ov}/\omega_{pi}$ of ~8 from Figure 25. This gives an estimation of the electrostatic signal width as $\omega_{ov}$ ~500 nm. However, when $A_{ov}/A_{pi} < 1$ there is a large variation of $\omega_{ov}/\omega_{pi}$ for small variation of $A_{es}$ (Fig. 25) which makes the estimation of the electrostatic signal particularly prone to large errors. This is *especially* true since $A_{es}$ is itself an estimation. Therefore, the electrostatic signal width, $\omega_{ov}$ is still an uncertain quantity in this material system.

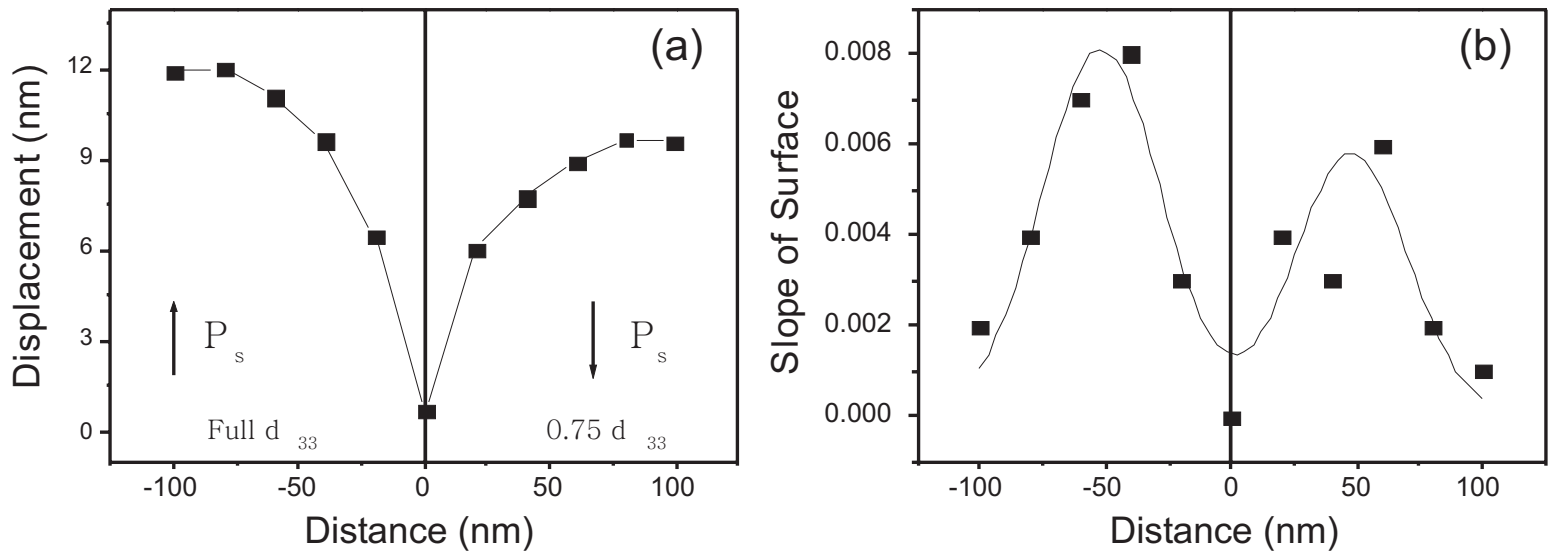

**Figure 26: FEM simulations of a domain wall with the $d_{33}$ coefficient of the right side of a 180$^o$ domain wall (at *x*=0) reduced to 75% of the full value on the left side. Shown in (a) is the vertical signal and in (b) the lateral signal 90$^o$ to the wall.**

Finally, the issue of the asymmetry in the PFM images will be examined. Any asymmetry in the electrostatic distribution across a wall could give rise to asymmetry in the vertical signal in a way discussed above. Measurements of the electrostatic distribution should be performed using non-contact methods, but the inherent long range nature of these measurements might not provide the spatial resolution needed to resolve the issue.[14, 40, 68, 69]

Another consideration is that the asymmetry could be due to changes in the material properties in the area of the domain wall. These highly stressed and distorted regions around the domain wall could have different physical properties from the bulk values. It has been shown by scanning-nonlinear-dielectric microscopy in periodically poled lithium niobate that very strong residual stresses or electric fields remain in the crystal that reduce the nonlinear dielectric constant in the region of the wall.[70] The asymmetry could be explained by a change in the wall region of any of the physical coefficients important to this measurement: the dielectric, piezoelectric, or elastic constants.

As an exaggerated example, FEM simulations were performed which arbitrarily reduced the $d_{33}$ coefficient on one side of the domain wall to 75% of the other side. The simulated results are shown in Figure 26(a). It shows that the vertical signal has some asymmetry because the right side of the domain wall does not expand as much as the left, as one would expect. Similar results can be drawn for the lateral signal in Figure 26(b) as well. This step-like large reduction of $d_{33}$ across a domain wall is perhaps a less likely scenario than a more realistic gradient of the value of $d_{33}$ across the wall. Such FEM calculations are more difficult with present commercial codes and require further work. Measurements made using the PFM setup give the same amplitude of the oscillation in an up and a down domain when measured far from the domain wall (>100 μm), which indicates that any changes must be in a highly localized region around the domain wall. The piezoelectric $d_{33}$ coefficient was chosen in this study for modeling simplicity, but modification of other piezoelectric coefficients, as well as the dielectric or the elastic constants are also possibilities.

## VI: CONCLUSIONS

The local piezoelectric response at a single ferroelectric 180° domain wall is measured in congruent and near-stoichiometric $LiNbO_3$ single crystals. Unexpected asymmetry in piezoresponse across the wall was observed, which is found to correlate to the crystal stoichiometry. The measured electromechanical interaction widths in congruent crystals are wider than in the near-stoichiometric values: for the vertical signal, $\omega_o$=140 nm compared to 113 nm, and for the lateral signal, 211 nm compared to 181 nm. Finite element modeling of the electromechanical response of the domain wall shows excellent qualitative agreement with experimental images for near-stoichiometric compositions. The amplitude of oscillation in vertical piezoresponse mode also showed an excellent agreement between modeling (13.4 nm) as compared to the measured (20-30 nm) values. Detailed analysis shows that the PFM resolution of a single antiparallel wall is determined both by intrinsic electromechanical width as well as tip size.

We acknowledge useful discussions with Dr. S. Kalinin, Dr. Gruverman, and Prof. D. Bonnell. This work was supported by National Science Foundation grant numbers DMR-9984691, DMR-0103354, and DMR-0349632.